\documentclass[fleqn,usenatbib]{mnras}

\usepackage{newtxtext,newtxmath}

\usepackage[T1]{fontenc}

\DeclareRobustCommand{\VAN}[3]{#2}
\let\VANthebibliography\thebibliography
\def\thebibliography{\DeclareRobustCommand{\VAN}[3]{##3}\VANthebibliography}

\usepackage{graphicx, float, hyperref, tabularx, physics, ulem, multirow}	
\usepackage{amsmath}	
\usepackage{amssymb}	

\usepackage{threeparttable} 

\newcommand{\gx}{{GX~339$-$4}}
\newcommand{\grs}{{GRS~1915+105}}

\newcommand{\rxte}{{\it{RXTE}}}

\title[The bump in \gx]{A systematic study of the high-frequency bump in the black-hole low-mass X-ray binary \gx}

\author[Y.\ Zhang et al.]{
Yuexin Zhang,$^{1,2}$\thanks{E-mail: yzhang@astro.rug.nl; yuexin.zhang@cfa.harvard.edu}
Mariano M\'{e}ndez,$^{1}$\thanks{E-mail: mariano@astro.rug.nl}
Sara E.\ Motta,$^{3}$
Andrzej A.\ Zdziarski,$^{4}$
\newauthor
Gr\'{e}goire Marcel,$^{5}$
Federico Garc\'{i}a,$^{6}$
Diego Altamirano,$^{7}$
Tomaso M.\ Belloni (deceased),$^{3}$
\newauthor
Liang Zhang,$^{8}$
Thimo Timmermans,$^{1}$
and Guobao Zhang$^{9,10}$
\\
$^{1}$ Kapteyn Astronomical Institute, University of Groningen, P.O.\ BOX 800, 9700 AV Groningen, The Netherlands\\
$^{2}$ Center for Astrophysics, Harvard \& Smithsonian, 60 Garden St, Cambridge, MA 02138, USA\\
$^{3}$ Istituto Nazionale di Astrofisica, Osservatorio Astronomico di Brera, via E.\ Bianchi 46, I-23807 Merate (LC), Italy\\
$^{4}$ Nicolaus Copernicus Astronomical Center, Polish Academy of Sciences, Bartycka 18, PL-00-716 Warszawa, Poland\\
$^{5}$ Institute of Astronomy, University of Cambridge, Madingley Road, Cambridge CB3 0HA, UK\\
$^{6}$ Instituto Argentino de Radioastronom\'{i}a (CCT La Plata, CONICET; CICPBA; UNLP), C.C.5, (1894) Villa Elisa, Buenos Aires, Argentina\\
$^{7}$ School of Physics and Astronomy, University of Southampton, Southampton SO17 1BJ, UK\\
$^{8}$ Key Laboratory of Particle Astrophysics, Institute of High Energy Physics, Chinese Academy of Sciences, Beijing 100049, People's Republic of China\\
$^{9}$ Yunnan Observatories, Chinese Academy of Sciences, Kunming 650216, People's Republic of China\\
$^{10}$ Key Laboratory for the Structure and Evolution of Celestial Objects, CAS, Kunming 650216, People's Republic of China
}

\date{Accepted XXX. Received YYY; in original form ZZZ}

\begin{document}
\label{firstpage}
\pagerange{\pageref{firstpage}--\pageref{lastpage}}
\maketitle

\begin{abstract}
The high-frequency bump, characterized by a frequency exceeding $\sim$30~Hz, represents a seldom-explored time-variability feature in the power density spectrum (PDS) of black-hole X-ray binaries. In the 2002, 2004, 2007 and 2010 outbursts of \gx, the bump has been occasionally observed in conjunction with type-C quasi-periodic oscillations (QPOs). We systematically study the properties of the bump during these four outbursts observed by \textit{Rossi X-ray Timing Explorer} (\rxte) in the 2--60~keV bands and detect the bump in 39 observations. While the frequencies of the type-C QPOs are in the range of $\sim$0.1--9~Hz, the root-mean-square (rms) amplitude of the bump shows an evolution in the hardness ratio versus the type-C QPO frequency plot. By comparing the rms amplitude of the bump with the corona temperature and simultaneous radio jet flux of the source, as previously studied in GRS 1915+105, we establish that in the hard state of \gx, the bump is always strong, with the measurements of the rms amplitude in the range of 4--10\%. At the same time, the corona temperature is high and the radio flux is low. These findings indicate that, using the bump as a proxy, the majority of the accretion energy is directed towards the hot corona rather than being channeled into the radio jet.
We discuss this phenomenon in terms of an inefficient energy transfer mechanism between the corona and jet in \gx.
\end{abstract}

\begin{keywords}
accretion, accretion discs -- stars: individual: \gx\ -- stars: black holes -- X-rays: binaries
\end{keywords}



\section{Introduction}\label{sec:intro}

Discovered as a variable X-ray source in 1973~\citep{1973ApJ...184L..67M}, \gx\ is a typical Galactic black-hole low-mass X-ray binary~\citep[BH LMXB;][]{2016A&A...587A..61C,2022arXiv220610053B}, showing regular luminous X-ray outburst every 2--3 years with a duty cycle of $\sim$30\%~\citep[e.g.][]{2015ApJ...805...87Y,2021MNRAS.507.5507A,2023MNRAS.tmp..760Y}. \gx\ possibly has a mass ratio between the donor star and the compact object of $0.18\pm 0.05$, a BH mass of 2.3--9.5~M$_{\odot}$, a binary orbital inclination of 37--78$^{\circ}$, and it is at a distance of $>5$~kpc~\citep{2004ApJ...609..317H,2017ApJ...846..132H,2019MNRAS.488.1026Z}. Broadband X-ray spectral analysis shows that the inner accretion disk has an inclination angle of approximately 30$^{\circ}$~\citep{2016ApJ...821L...6P} and the black hole has a dimensionless spin parameter of approximately 0.9~\citep{2008ApJ...679L.113M,2016ApJ...821L...6P}.

Most BH LMXBs, including \gx, are typical X-ray transients that during outburst from months to years trace an anti-clockwise `q' path in the hardness-intensity diagram~\citep[HID;][]{2005A&A...440..207B,2005Ap&SS.300..107H}. The relative proportion of the soft thermal disk and the hard corona component to the X-ray flux determines the spectral state of the source~\citep[for a review, see][]{2010LNP...794...17G}. In the HID, there are several well-defined spectral states~\citep{2005A&A...440..207B}. The source spends most of the time in the X-ray quiescent state before it goes into outburst~\citep[see][for a review]{2006ARA&A..44...49R}. At the start of an outburst, the X-ray luminosity of the source increases by several orders of magnitude compared to the quiescent state and the source enters the low-hard state (LHS), which is dominated by the hard corona component~\citep[e.g.][]{2018ApJ...855...61W}. As the outburst continues, the emission of the soft thermal disk gradually increases and the source undergoes a state transition from the LHS to the hard-intermediate state (HIMS), soft-intermediate state (SIMS), high soft state (HSS) and sometimes an anomalous ultra-luminous state~\citep[e.g.][]{1997ApJ...479..926M,2004ApJ...601..428K,2012MNRAS.427..595M,2023MNRAS.520.5144Z}. In the end of an outburst, the source goes back to the LHS and finally to quiescence. An outburst usually lasts from weeks to months~\citep{2006ARA&A..44...49R}, and sometimes the X-ray transients, including \gx, become active but never transition into the HSS, in what is known as a failed-transition outburst~\citep[][and references therein]{2021MNRAS.507.5507A}.

Radio jets are prominent ejections as the spectral state of an X-ray transient changes~\citep[see][for a review]{2005Ap&SS.300....1F}. In the LHS and the HIMS, a compact radio jet as a result of synchrotron emission can be observed, with the radio flux correlated with the X-ray flux~\citep[e.g.][]{2001ApJ...554...43C,2003MNRAS.344...60G}. During the state transition from the HIMS to SIMS, this compact radio jet is quenched, and in the SIMS a powerful transient jet sometimes with spatially-resolved ejecta appears~\citep[e.g.][]{2002Sci...298..196C,2004MNRAS.355.1105F}. The transient jet disappears when the X-ray transient enters the HSS, while a compact jet reappears as the source returns to the hard state at the end of an outburst~\citep[e.g.][]{2011ApJ...739L..19R,2013MNRAS.428.2500C}.

Apart from the long-term evolution of black-hole X-ray binaries (BHXRBs), short time variability is a powerful tool to study the accretion flow from the X-ray to the radio bands~\citep[e.g.][]{2003MNRAS.345..292H,2019MNRAS.484.2987T,2019NewAR..8501524I,2021MNRAS.503..614V}. By applying Fourier techniques to the X-ray light curves, the time variability components in the frequency domain present in the power density spectrum (PDS) can generally be decomposed into narrow peaks called quasi-periodic oscillations~\citep[QPOs;][]{2019NewAR..8501524I} and broadband noise (BBN) components~\citep{2000MNRAS.318..361N,2002ApJ...572..392B}. The low-frequency QPOs (LFQPOs) are classified into three types, types A, B, and C, basically depending on the QPO frequency ($\nu$), the quality factor ($\nu$/FWHM), the fractional root-mean-square amplitude (hereafter rms) and the strength of the BBN~\citep{2004A&A...426..587C,2005ApJ...629..403C}. Type-C QPOs are common in the LHS and HIMS, while type-A/B QPOs can appear in the HSS/SIMS~\citep{2016AN....337..398M}. High-frequency QPOs (HFQPOs), with central frequency in the range of $\sim$30~Hz to hundreds of Hz, are less frequent than LFQPOs and have only been detected in a handful of BHXRBs~\citep[e.g.][and references therein]{2012MNRAS.426.1701B,2013MNRAS.435.2132M,2022MNRAS.517.1469M}. In PDS with QPOs, the BBN components have characteristic frequencies that are strongly correlated with the QPO frequency, indicating the same dynamical origin of the BBN and the QPOs~\citep{1999ApJ...514..939W,1999ApJ...520..262P,2002ApJ...572..392B}.

In some BHXRBs, an upper high-frequency bump (hereafter ``the bump'') at frequencies higher than $\sim$30~Hz has been detected but studied less often than the LFQPOs~\citep[e.g.][]{2001ApJ...558..276T,2002ApJ...572..392B,2003A&A...407.1039P}. The bump was first associated with the X-ray corona and denoted by $L_{u}$ in~\citet{2002ApJ...572..392B}. Recently, \citet{2022MNRAS.514.2891Z} performed a systematic study of the bump in \grs\ when the type-C QPOs appear in this source from 1996 to 2012. In the so called `plateau' state~\citep{1999MNRAS.304..865F}, \grs\ shows continuous radio flares at 15~GHz, as seen with Ryle telescope~\citep{1997MNRAS.292..925P,2022NatAs...6..577M}. \citet{2022NatAs...6..577M} and~\citet{2022MNRAS.514.2891Z} suggested that the bump should originate from the X-ray corona, and that the bump and the 67-Hz HFQPO in \grs~\citep[e.g.][]{2012MNRAS.426.1701B} represent the same variability component but the characteristics of the corona determine the coherence of this component. (See~\citealt{2023arXiv230700867M} for a quantitative explanation.) More importantly, an anti-correlation between the rms of the bump and the radio flux, and the correlation between the rms of the bump and the corona temperature of the source, suggest that the corona functions as the energy reservoir for the radio jet~\citep{2022MNRAS.514.2891Z}. However, \grs\ is peculiar since it did not follow the `q' path in the HID of typical black-hole X-ray transients~\citep{2004ARA&A..42..317F}, making the findings in~\citet{2022MNRAS.514.2891Z} quite special at that stage. 

In some  black-hole X-ray transients the characteristic frequency of the bump was reported (e.g., XTE~J1650$-$500, \citealt{2003ApJ...586..419K}, XTE~J1118+480, \citealt{2002ApJ...572..392B}, MAXI~J1348$-$630, \citealt{2022MNRAS.514.2839A}), confirming the correlation between the frequency of the bump and that of the LFQPOs. Furthermore, the frequency of the bump was used to infer the spin of the black-hole transients, GRO~J655$-$40~\citep{2014MNRAS.437.2554M}, XTE~J1550$-$564~\citep{2014MNRAS.439L..65M}, MAXI~J1820+070~\citep{2021MNRAS.508.3104B} and XTE~J1859+226~\citep{2022MNRAS.517.1469M}. While these papers focus on the dynamical properties of the bump, the evolution of the radiative properties of the bump have been rarely studied.
In \gx, the bump was fitted in a handful of \textit{Rossi X-ray Timing Explorer}~\citep[\textit{RXTE};][]{1993A&AS...97..355B} observations~\citep{2000MNRAS.318..361N,2002ApJ...572..392B}. \citet{2005A&A...440..207B} studied only the 2002 outburst of \gx\ and found that the rms of the bump is always around 2--3\%.

Here, we report the first systematic study of the bump in the black-hole X-ray transient \gx\ using the archival data of \rxte. This paper is organized as follows: In section~\ref{sec:analysis} we describe the \rxte\ data reduction. We generate the PDS and the time-averaged energy spectra and the corresponding background spectra and response matrices. We perform a spectral-timing analysis of these observations and obtain the radio flux density using data from~\citet{2013MNRAS.428.2500C}. In section~\ref{sec:results} we present the results, including examples of the PDS, the rms of the bump in the hardness ratio (HR) versus QPO frequency diagram, the QPO frequency evolution accompanied by the compact jet flux density, and correlation of the rms of the bump with the radio flux and the X-ray corona temperature. In section~\ref{sec:discussion} we discuss our results and compare them with our previous studies of the bump~\citep{2022MNRAS.514.2891Z} in the source \grs.

\section{Observations and data analysis}\label{sec:analysis}

The archival data of \gx\ contain $\sim$1000 observations with \rxte\ during its lifespan from 1996 to 2012. We extract the light curve of the source, one point per observation, and subtract the background count rate. The count rate is then obtained separately for the 7--13 keV (hard) band and the 2--7 keV (soft) band, both in units of the Crab nebula, as described in~\citet{2008ApJ...685..436A}. We compute the hardness ratio by dividing the count rate in the hard band by that in the soft band.

We generate the energy spectra of \gx\ from the \rxte/Proportional Counter Array~\citep[PCA;][]{2006ApJS..163..401J} and High Energy X-ray Timing Experiment~\citep[HEXTE;][]{1998ApJ...496..538R} within HEASOFT V6.32 following the standard pipelines given in the \rxte\ cookbook. We extract PCA spectra from the Proportional Counter Unit 2 (PCU2) which is the most accurately calibrated detector among the five PCUs, and HEXTE spectra from the Cluster B only. We generate the associated response matrices and add a systematic uncertainty of 0.6\% to the PCA spectra to account for calibration uncertainties. We create the background spectra by applying the background model suitable for the brightness level of \gx.

\rxte\ observed four outbursts of \gx\ in 2002, 2004, 2007 and 2010~\citep[e.g.][]{2015ApJ...805...87Y,2016AN....337..435C}. Among the observations of these outbursts we only investigate further the 91 observations associated with the type-C QPOs in the LHS and HIMS as reported in~\citet{2020A&A...640A..18M}.

\subsection{Timing analysis}

For each observation we take a time segment of a length of 32~s using all the PCA channels, perform fast Fourier transformation (FFT), and average all the FFT segments to obtain a single PDS per observation. The time resolution is always at least 1/512~s such that the Nyquist frequency is 256~Hz. We subtract the Poisson noise level~\citep{1995ApJ...449..930Z} and normalize the PDS to units of rms$^2$ per Hz~\citep{1990A&A...230..103B}. We do not consider the background rate to convert the PDS to rms units since it is negligible compared to the source rate in our observations. We apply logarithmic rebinning to the frequency of the PDS so that the size of the frequency bins increases by $\exp(1/100)$ compared to the previous frequency bin.

We fit the PDS in the frequency range of 0.03125--256~Hz in XSPEC version 12.13.1~\citep{1996ASPC..101...17A}. We use three Lorentzian functions to represent the QPO and two BBN components (see~\citealt{2000MNRAS.318..361N} or~\citealt{2002ApJ...572..392B} for the definition of the Lorentzian function). The central frequency of one of the two Lorentzians that fit the broadband noise is fixed at 0. If a QPO harmonic and/or subharmonic is present in the PDS, we add one or two more Lorentzians to fit them. In some cases, the power spectra are not satisfactorily fitted by all these Lorentzian functions, for example, due to the shift of the peaks within one observation and a third harmonic. To fit these residuals, we add extra Lorentzian functions. We have also noticed that in some PDS, there is enhanced power, due to the high-frequency bump in the frequency range $\sim$20--200~Hz~\citep{2001ApJ...558..276T,2005A&A...440..207B,2022MNRAS.514.2891Z}. Therefore, we use one additional Lorentzian centered at 0~Hz to fit the bump.

We test the significance of the bump by computing the width and normalization of the Lorentzian function that fits it. If the bump is badly constrained (the 1-$\sigma$ boundary of the width of the bump falls $< 20$~Hz and/or $> 180$~Hz), we fix the width of the Lorentzian at 70~Hz and calculate the 95\% upper limits of its normalization. This normalization is consistent with the value of the case where the width of the bump is badly constrained. We also check that there is no bump with width $\lesssim$ 20~Hz. If the bump has a good constraint on its width but its normalization is less than 3-$\sigma$ significant, we also calculate the 95\% upper limit of its normalization. Finally, we convert the normalization of the bump to rms units.

\subsection{Spectral analysis}

We analyze the energy spectra of PCA in the $\sim$2--25~keV band. We use the XSPEC components \texttt{diskbb}~\citep{1984PASJ...36..741M} and \texttt{nthcomp}~\citep{1996MNRAS.283..193Z,1999MNRAS.309..561Z} to describe the disk blackbody emission and direct corona emission, respectively. So the initial model used to fit the time-averaged energy spectra is \texttt{tbabs*(diskbb+nthcomp)} where \texttt{tbabs} accounts for the interstellar absorption towards the source. The temperature of the seed photons in \texttt{nthcomp} is linked to the inner disk temperature in \texttt{diskbb}. For \texttt{tbabs} we use the cross-section of~\citet{1996ApJ...465..487V} and the solar abundance of~\citet{2000ApJ...542..914W}. Following~\citet{1997ApJ...479..926M}, we fix the Galactic absorption column density, $N_{\text{H}}$, at $5\times 10^{21}$~cm$^{-2}$.

After the initial fitting of the PCA data, we find clear residuals at $\sim$6.4~keV and $\sim$20~keV, probably due to a relativistic broadened iron line and a Compton hump, respectively~\citep{1988ApJ...335...57L,1989MNRAS.238..729F}. We therefore add a \texttt{relxillCp} component~\citep{2014ApJ...782...76G} such that the final model is \texttt{tbabs*(diskbb+nthcomp+relxillCp)}. Given the limited resolution of the PCA energy spectra and the complexity of the reflection model, we fix several parameters in \texttt{relxillCp} following~\citep{2021MNRAS.508..287S}: We fix the corona emissivity indices to index1 = index2 = 3~\citep{1989MNRAS.238..729F}, the disk inclination angle, $i$, to 30$^{\circ}$~\citep{2016ApJ...821L...6P}, the dimensionless spin parameter, $a*$, to 0.93~\citep{2008ApJ...679L.113M,2016ApJ...821L...6P}, and the iron abundance, $A_{\text{Fe}}$, to 6.6 times of the solar abundance~\citep{2016ApJ...821L...6P,2018ApJ...855...61W}, which is too high to be real, but likely an artefact of the model~\citep{2018ASPC..515..282G}. All the fixed parameters are basically consistent with the results from the most recent detailed spectral analysis of \gx~\citep{2022MNRAS.513.4308L,2023ApJ...950....5L}. In the reflection model, the inner radius of the accretion disk, $R_{\text{in}}$, and the disk ionization parameter, $\log\xi$, are fitted freely. The corona temperature, $kT_{\text{e}}$, in \texttt{relxillCp} is linked to that in \texttt{nthcomp}.

In some observations the values of $kT_{\text{e}}$ from fitting the energy spectra of PCA in $\sim$2--25~keV band are not constrained, we combine the HEXTE data, whenever available in these observations, in the 20--200~keV band with the PCA data in order to obtain better constraints on $kT_{\text{e}}$.

\subsection{Radio data}

We use the 8.6 or 9.0~GHz radio data from the Australia Telescope Compact Array (ATCA) as reported in~\citet{2013MNRAS.428.2500C}. To compare the radio flux density of \gx\ with that of \grs\ in our previous study~\citep{2022NatAs...6..577M,2022MNRAS.514.2891Z,2022MNRAS.513.4196G}, we obtain the radio flux density of \gx\ by placing it at the distance of \grs. We consider that the distance to \gx~\citep{2017ApJ...846..132H,2019MNRAS.488.1026Z} and \grs~\citep{2014MNRAS.444.1113Z,2014ApJ...796....2R} are both $\sim$9~kpc, and estimate the uncertainties of the radio flux density of \gx\ based on the distance uncertainty from the most recent study by~\citep{2019MNRAS.488.1026Z}. We then convert the flux density of \gx\ from 8.6~GHz or 9~GHz to 15~GHz assuming a radio spectral index of $\alpha=0.4$~\citep{2013MNRAS.428.2500C,2018MNRAS.481.4513I}.

We notice that there are only a few simultaneous ($< 1$ day) radio and X-ray observations. So we perform a linear interpolation of the radio data to obtain values at the times of the X-ray observations. We interpolate the radio data over gaps of less than 10 days to have interpolated radio data on a 1-day interval, and match the interpolated radio data with the X-ray observations.

\section{Results}\label{sec:results}

\subsection{Power density spectrum and the bump}\label{subsec:pds}

\begin{figure*}
    \includegraphics[width=0.99\textwidth]{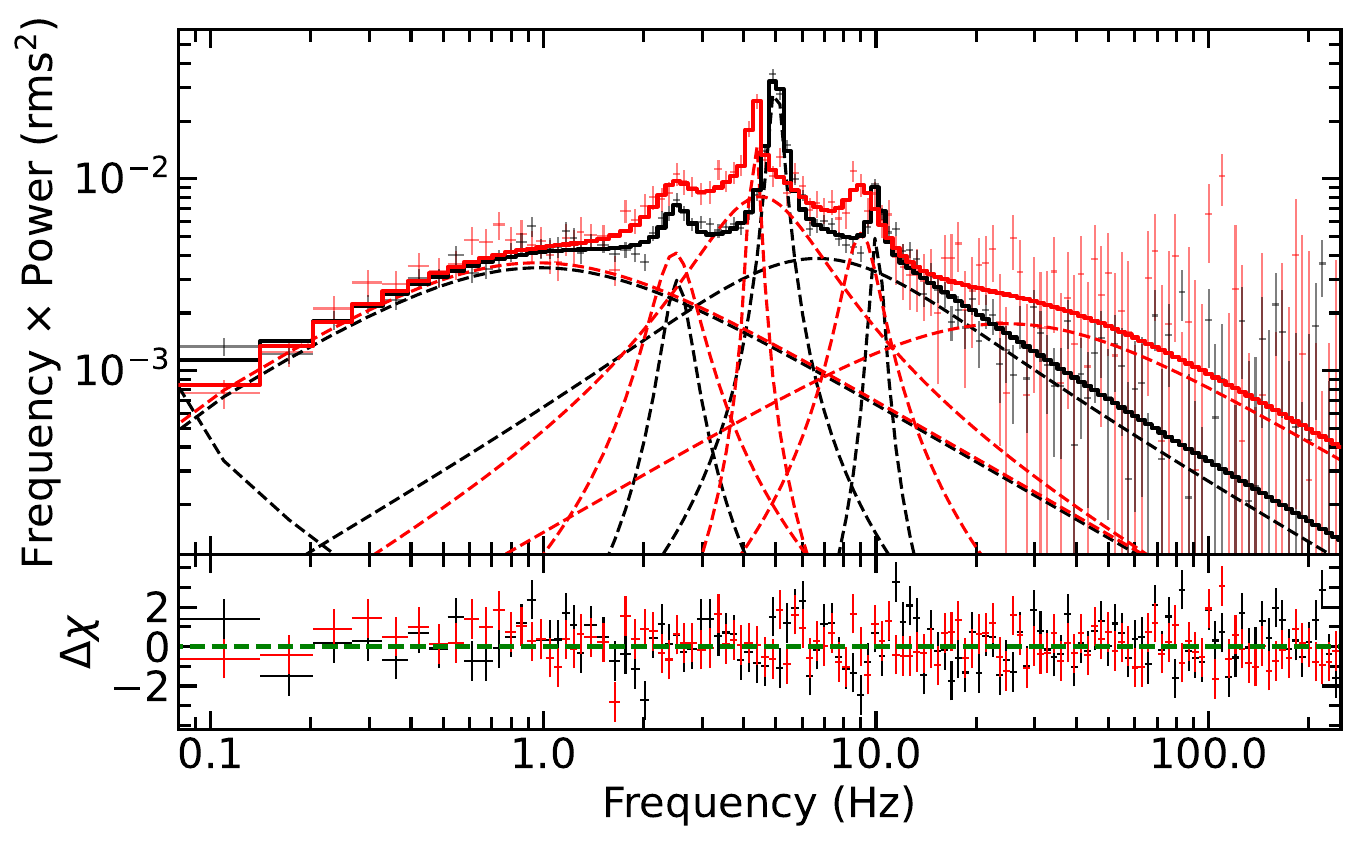}
    \caption{PDS of two observations of \gx\ with the type-C QPOs. The black and red points correspond to observation IDs 92428-01-04-03 and 60705-01-70-00, respectively, both of which are in the hard state, in the rising part of the HID. Top panel: Points with error bars are data, solid lines represent the best-fitting models, while dashed lines represent the individual Lorentzian functions that fit the data. Bottom panel: The residuals with respect to the best-fitting model.}
    \label{fig:pds}
\end{figure*}

Fig.~\ref{fig:pds} shows two examples out of the 91 \rxte/PCA PDS of \gx\ with type-C QPOs, corresponding to observation IDs 92428-01-04-03 (black; 2007 outburst) and 60705-01-70-00 (red; 2004 outburst). The black PDS corresponds to the case in which the bump is not significantly detected, while the red PDS has a significant bump. In both cases, the type-C QPO fundamental is fitted with a narrow Lorentzian at 5~Hz (black) and at 4.3~Hz (red). QPO harmonic and subharmonic appear at approximately 0.5 and 2 times the QPO fundamental frequency, respectively. Two extra Lorentzians are required to fit the BBN in both cases.
The 95\% upper limit to the rms amplitude of the zero-centered Lorentzian function that fits the bump is 3.9\% in black PDS while the rms amplitude of the bump in the red PDS is of $7.5\pm 0.7\%$. Overall, the QPO frequency is in the range of $\sim$0.1--9~Hz, while the HR is between $\sim$0.2 and 1.4 in Crab units. The width of the bump that is significantly detected shows a correlation with the QPO fundamental frequencies, consistent with the correlation between the $L_{u}$ and $L_{\text{QPO}}$ in BHXRBs as shown by, e.g.,~\citet{2002ApJ...572..392B}.

\subsection{The rms of the bump}\label{subsec:rms}


\begin{figure*}
    \includegraphics[width=0.99\textwidth]{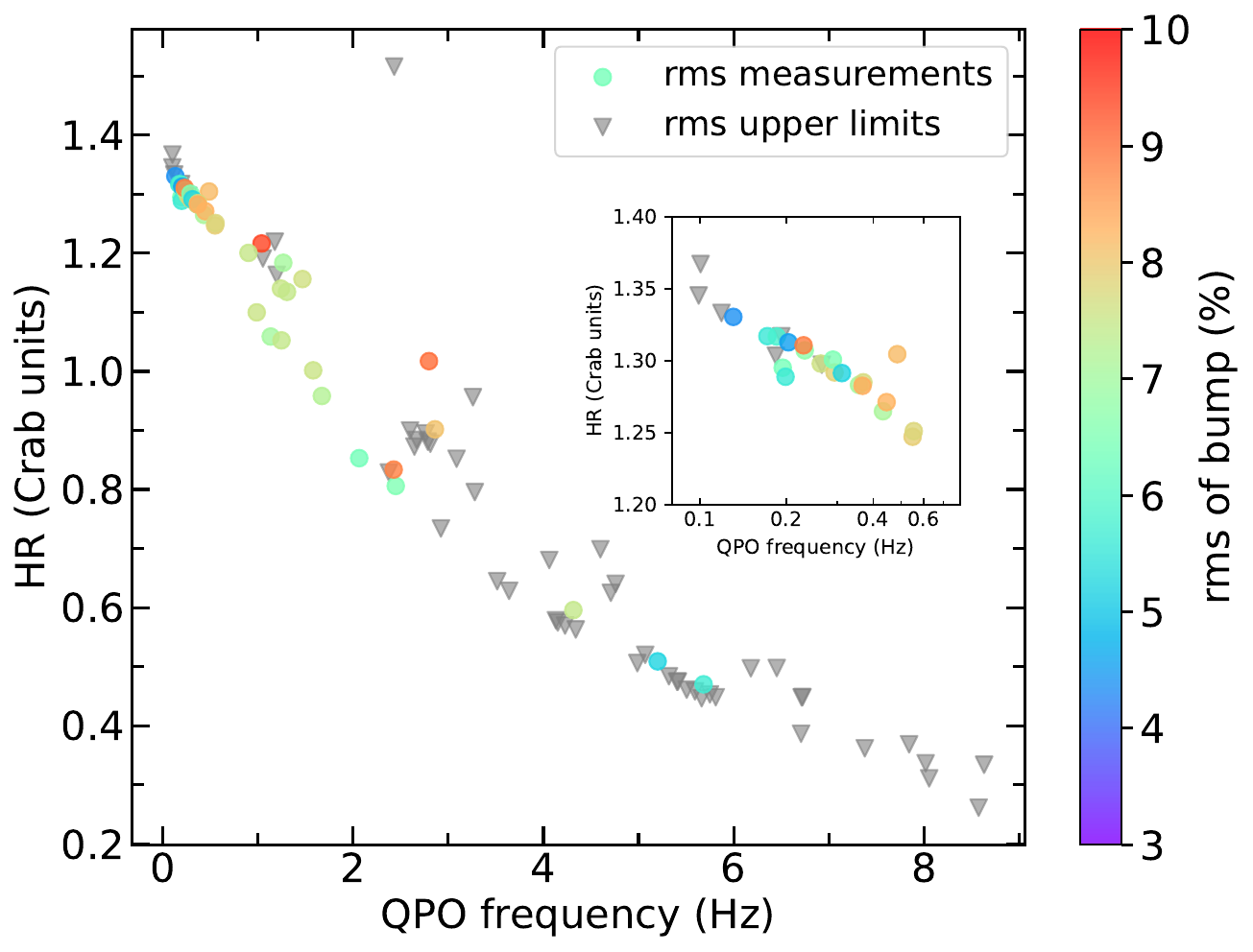}
    \caption{Hardness ratio versus type-C QPO frequency plot for the 91 observations of \gx. The color of the points indicate the rms amplitude of the bump. The colored circles indicate the significant measurements, while the grey triangles are upper limits. There is a strong bump when the HR is high, while no bump when the HR is low. The small panel inside shows the rms amplitude of the bump when the QPO frequency is near 0.1~Hz.}
    \label{fig:rms}
\end{figure*}


In Fig.~\ref{fig:rms}, we plot the HR of the source versus the type-C QPO frequency using the 91 \rxte\ observations of \gx\ in the LHS and HIMS. Error bars of both HR and QPO frequency are within the size of the data points. The HR and QPO frequency follow a clear anti-correlation, i.e.\ the HR generally decreases as the QPO frequency increases. The data points are not as scattered around the anti-correlation as those in Fig.~1 in~\citet{2022MNRAS.514.2891Z} for \grs, possibly due to the fact that there are far less observations of \gx\ (91 observations) than of \grs\ (410 observations).

The rms amplitude of the bump in the HR versus QPO frequency plot, Fig.~\ref{fig:rms}, is between $\sim$4\% and $\sim$10\%. The circles are the observations with significant bumps, while the triangles are the observations with upper limits. The rms measurements have an average error of $\pm 0.7\%$. As the source softens, the rms amplitude of the bump decreases. When the HR is around 1.3 and the QPO frequency is near 0.1~Hz, the rms amplitude of the bump is $\sim$4--9\%, while when the HR is $\sim$0.2 and the QPO frequency increases from 0.1~Hz to 8~Hz, the rms of the bump generally decreases from $\sim$9\% to $< 4\%$. In the relatively softer observations, with HR $\lesssim 0.8$ and QPO frequency $\gtrsim 3$~Hz, the bump is generally not detected significantly.

Note that there is an outlier point when the QPO frequency is $\sim$2~Hz and the HR is $\sim$1.5. This observation, in the LHS, is harder than any other observation that we analyze. In this observation the rms of the bump is also not significant, indicated by the down-pointing triangle.

\begin{figure*}
    \includegraphics[width=0.99\textwidth]{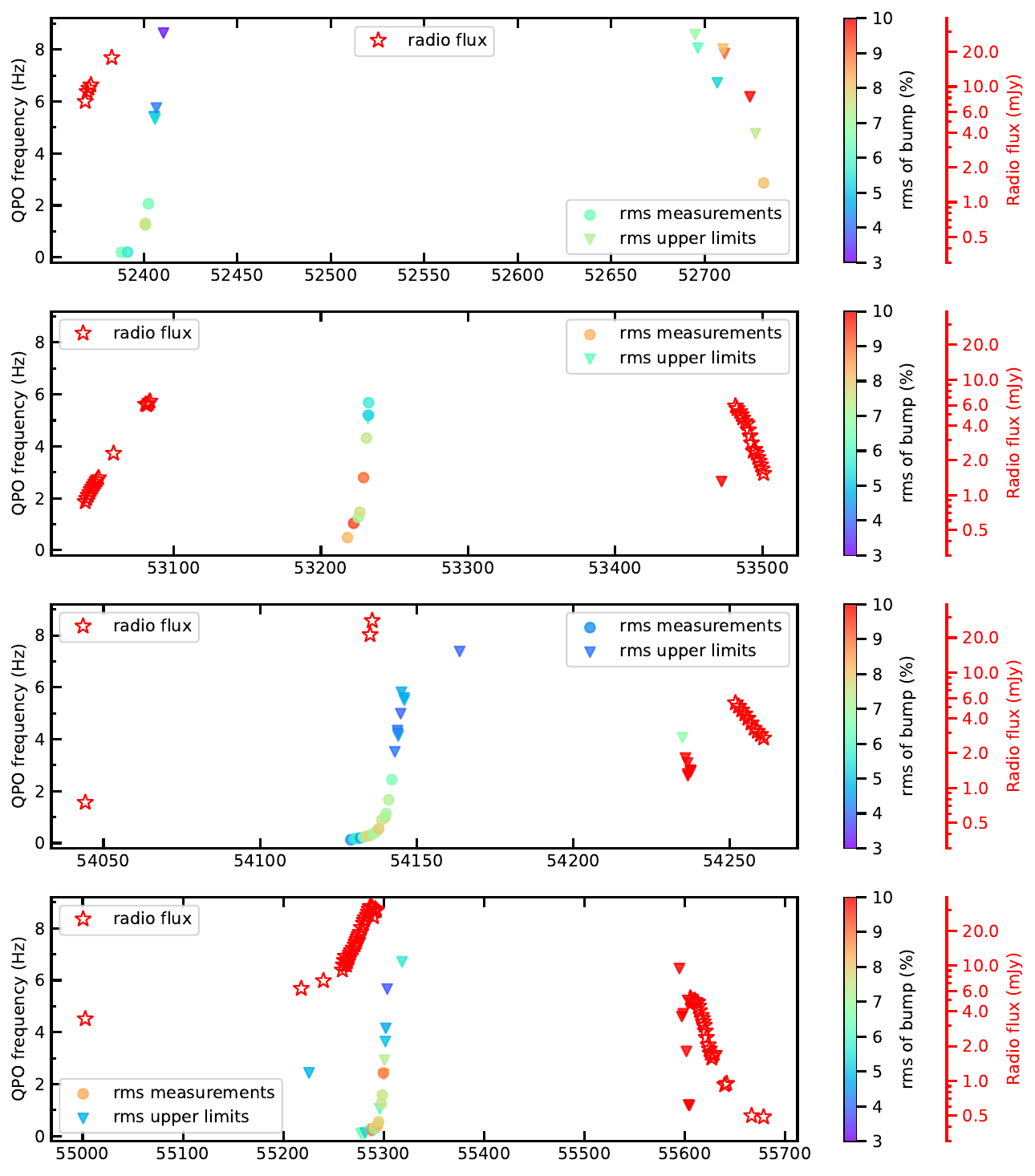}
    \caption{Temporal evolution of the QPO frequency, rms of bump, and the radio luminosity of \gx\ are shown in four panels from top to bottom, representing the 2002, 2004, 2007, 2010 outbursts, respectively. The circles and triangles represent the QPO frequency. The QPO frequency is denoted by circles and triangles, with colors indicating measurements and upper limits of the rms of the bump, respectively. Note that in the decaying phase, the red triangles represent the upper limits of the rms that exceed 10\%. The 8.5 or 9~GHz radio luminosity of \gx\ from ATCA observations is denoted by red empty stars. Because of the limited number of simultaneous X-ray and radio observations, most of the X-ray observations with a type-C QPO lack simultaneous radio observations.}
    \label{fig:rms_radio}
\end{figure*}

We plot the temporal evolution of the type-C QPO frequency for the 2002, 2004, 2007 and 2010 outbursts of \gx\ in Fig.~\ref{fig:rms_radio}. In these four outbursts, the type-C QPO frequency ranges from $\sim$0.1 to 8~Hz. In the left-hand side of each panel in Fig.~\ref{fig:rms_radio}, the QPO frequency increases as the source evolves from the LHS to the HIMS, while in the right-hand side of the panels the QPO frequency decreases as the source evolves back to LHS~\citep{2011MNRAS.418.2292M,2020A&A...640A..18M}. The gaps in these panels indicate the times that \gx\ was in the SIMS and HSS. In the rising phase of the outburst, as the QPO frequency increases, the rms amplitude of the bump generally increases from < 4\% to $\sim$9\% and then decreases to below 4\%. The bump is no longer significant when the QPO frequency is $\gtrsim 3$~Hz. In the decaying phase of the outbursts, when the QPO frequency decreases, the bump is mostly not significant and the upper limits cannot be judged easily from the plot and show no trend.

We also plot in Fig.~\ref{fig:rms_radio} the radio flux of \gx. In the 2002 outburst, the observed radio flux increases from 6~mJy to 20~mJy during MJD~52350--52400. In the 2004 outburst, there are several radio observations from MJD~53000 to 53100, with the radio flux increasing from 1~mJy to $\sim$6~mJy. Between MJD~53480 to 53500, the radio flux decreases from $\sim$6~mJy to 1~mJy, however, those radio observations are not simultaneous with \rxte\ observations in the X-ray band. In the 2007 outburst the radio observations are sparse, but an increasing trend of the radio flux from 0.5~mJy to $>20$~mJy from MJD~54050 to 54150, and a decreasing trend from 6~mJy to $\sim$2~mJy after MJD~54250 are still apparent in the data. In the 2010 outburst, from MJD~55000 to 55300 the radio flux increases from 5~mJy to $\sim$30~mJy, while from MJD~55600 to 55700, the radio flux decreases from approximately 6~mJy to 0.5~mJy. Most X-ray data with type-C QPOs have no simultaneous radio observations. It is clear that, especially in the rising phase of the outburst, the increase of the QPO frequency generally lags the increase of the jet flux density.

It is worth noting that from MJD~53200 to 53250 during the 2004 outburst, the rms amplitude of the bump increases up to 9\% and decreases to $\sim$6\% at the end of the rising phase. During the same outburst, the radio flux is in the range of 0.5--2~mJy. The range of both the rms amplitude of the bump and the radio flux is quite different from those in the 2002, 2007 and 2010 outbursts of \gx.

\subsection{The rms of the bump vs.\ spectral parameters and radio flux}

\begin{figure*}
    \includegraphics[width=0.99\textwidth]{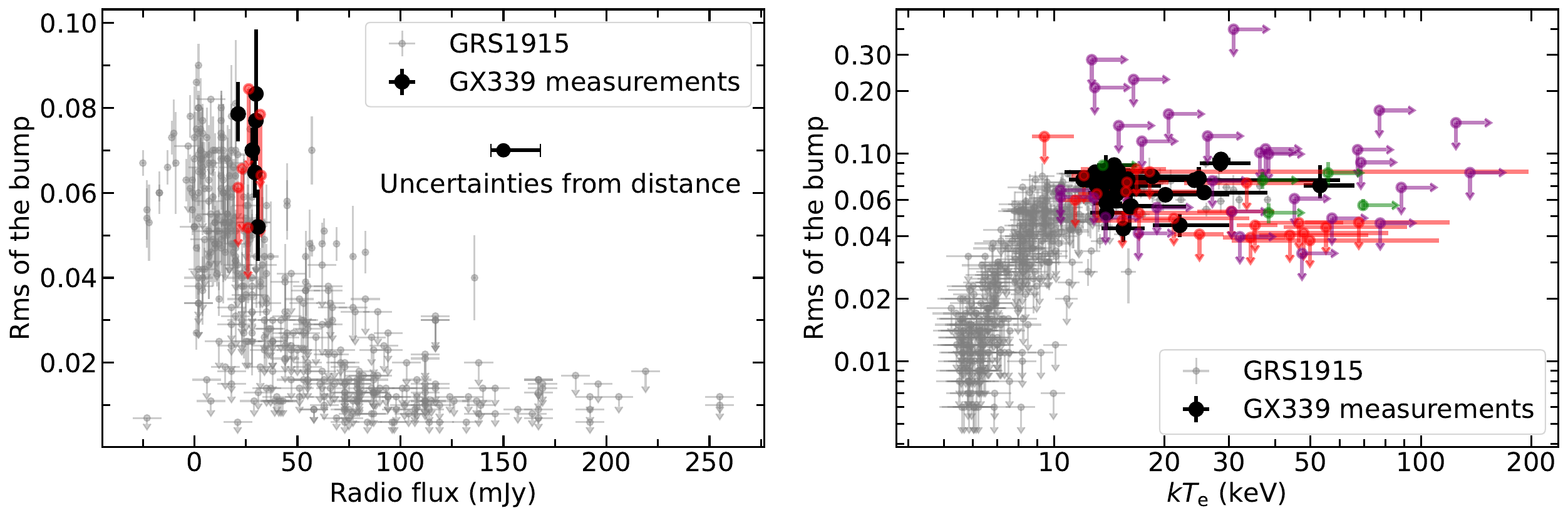}
    \caption{Left panel: The simultaneous X-ray and radio observations of \gx\ during the 2002, 2004, 2007 and 2010 outbursts. The black points indicate the measurements of the rms of the bump, while the red points indicate the upper limits of the rms of the bump. Right panel: The rms of the bump versus the corona temperature, $kT_{\text{e}}$. The black points represent the measurements of both the rms of the bump and $kT_{\text{e}}$; the red points represent the measurements of $kT_{\text{e}}$ but the upper limits of the rms of the bump; the green points represent the measurements of the rms of the bump but the lower limits of $kT_{\text{e}}$; the purple points represent the upper limits of both the rms of the bump and $kT_{\text{e}}$. In both panels, the grey points represent the measurements of \grs\ as reported in~\citet{2022MNRAS.514.2891Z}. Error bars indicate the 1-$\sigma$ error range.}
    \label{fig:correlation}
\end{figure*}

Fig.~\ref{fig:correlation} illustrates the measurements of the rms amplitude of the bump and the radio flux / corona temperature, $kT_{\text{e}}$, together with the previous measurements of \grs. The left panel of Fig.~\ref{fig:correlation} displays the X-ray and radio observations of \gx. The measurements and upper limits of the rms amplitude of the bump are plotted in black and red, respectively. When \gx\ is placed at the distance of \grs, the measurements of the radio flux are around 10~mJy, while the measurements and the upper limits of the rms of the bump are 5--9\% and $\lesssim$ 8\%, respectively. Despite there being only a handful of simultaneous X-ray and radio observations of \gx\ compared to the case of \grs~\citep{2022MNRAS.514.2891Z} where there is an anti-correlation between the rms amplitude of the bump and the radio flux, the measurements of \gx\ fall in the top portion of the measurements of \grs\ (grey points). The data of \gx\ in the left panel of Fig.~\ref{fig:correlation} show that the bump in \gx\ is quite strong, while the radio flux is low compared to the case of \grs.

We plot the rms amplitude of the bump versus the corona temperature, $kT_{\text{e}}$, in the right panel of Fig.~\ref{fig:correlation}. The black points represent the measurements of both the rms amplitude of the bump and $kT_{\text{e}}$, the red points represent the measurements of $kT_{\text{e}}$ and the upper limits of the rms amplitude of the bump, the green points represent the measurements of the rms amplitude of the bump and the lower limits of $kT_{\text{e}}$, and the purple points represent the upper limits of both the rms amplitude of the bump and $kT_{\text{e}}$. Although there are many upper and lower limits of either the rms amplitude of the bump and/or $kT_{\text{e}}$, in the case of \gx\ the measurements of both quantities are located in the top right part of the plot, overlapping with the data of \grs. The significant measurements of the bump have rms amplitudes between 4--10\%, while in the cases in which the bump is not significant the upper limits of the rms amplitude are generally $\lesssim 10\%$. Some upper limits of the rms amplitude, which only appear in the decaying phase of the outburst (Fig.~\ref{fig:rms_radio} and Fig.~\ref{fig:hid}), exceed 10\% and can be up to 30\%. The measurements of the corona temperature, $kT_{\text{e}}$, are in the range $\sim$15--80~keV, indicating a quite hot corona compared to the corona of \grs\ whose temperature can be down to 6~keV~\citep{2022MNRAS.514.2891Z}. The rms amplitude of the bump and $kT_{\text{e}}$ are correlated in \grs~\citep{2022MNRAS.514.2891Z}, indicating that both the corona is hotter and the bump is stronger. Compared to the data of \grs,  the rms amplitude of the bump and $kT_{\text{e}}$ are never weakened in the LHS and HIMS in \gx.

We find no correlation between the rms amplitude of the bump and the inner disk temperature, $kT_{\text{in}}$, since most fitting results show unexpected high $kT_{\text{in}}$ in the LHS and HIMS, consistent with the report from systematic spectral studies on \gx\ in~\citet{2016AN....337..435C} and~\citet{2021MNRAS.508..287S}.

\section{Discussion}\label{sec:discussion}

We have gone through all the \rxte\ archival data of \gx\ and further studied the 91 observations with type-C QPOs in the LHS and HIMS. We carried out the first systematic study of the high-frequency bump in a black hole X-ray transient, complementing our previous study of the bump in the source \grs. We fitted the bump in the PDS of \gx\ with a zero-centered Lorentzian function and found that it has a characteristic frequency in the range 20--200~Hz. Among the 91 observations, 39 show a significant bump and 52 show no bump, yielding only upper limits of its rms amplitude. As shown in Fig.~\ref{fig:rms}, the rms of the bump depends on the frequency of the type-C QPO and the HR. When the HR is between $\sim$0.8 and 1.3 and the QPO frequency is between $\sim$0.1~Hz and 2~Hz, the rms amplitude of the bump is $\sim$4--9\%, whereas when the HR decreases below 0.8 down to 0.2 and the QPO frequency increases from $\sim$2~Hz up to 8~Hz the bump is not detected with upper limits of the rms amplitude between 3\% and $\sim$10\%. In the case of \grs, we have found that the energy available in the system can be directed to power either the corona or the jet, depending on the strength of the bump and the radio flux and the temperature of the corona~\citep{2022NatAs...6..577M,2022MNRAS.513.4196G,2022MNRAS.514.2891Z}. Compared to our previous results in \grs, the radio flux of \gx\ is quite low, but the rms amplitude of the bump is high and the X-ray corona is hot, indicating that in \gx\ more energy in the system in the LHS and the HIMS is directed towards the X-ray corona than in the case of \grs. However, we recall that to estimate the actual distribution of the energy in the system requires proper modeling and a better understanding of both the radiative efficiencies and the variability features from the jets and the disk. This is beyond the scope of this work.

\subsection{On the detection of the bump}\label{subsec:bump}

\citet{2002ApJ...572..392B} has laid down the framework for fitting the lower high-frequency bump ($L_{\text{l}}$ in~\citealt{1999ApJ...520..262P}) and upper high-frequency bump ($L_{\text{u}}$ for the bump in this paper) in BHXRBs. As shown in~\citet{1999ApJ...520..262P}, the LFQPO frequency $L_{\text{LF}}$ in the $\sim$0.1--10~Hz range is correlated with the characteristic frequency of $L_{\text{l}}$ in the $\sim$1--100~Hz range. The correlation between $L_{\text{LF}}$ and $L_{\text{u}}$ also exists, with the characteristic frequency of $L_{\text{u}}$ in the $\sim$20--200~Hz range~\citep{2002ApJ...572..392B}. The bump $L_{\text{u}}$ has been reported in some BHXRBs, e.g. \grs~\citep{2001ApJ...558..276T}, XTE~J1118+480~\citep{2002ApJ...572..392B}, Cygnus~X-1~\citep{2003A&A...407.1039P}, \gx~\citep{2000MNRAS.318..361N}, XTE~J1650$-$500~\citep{2003ApJ...586..419K}, GRO~J655$-$40~\citep{2014MNRAS.437.2554M}, XTE~J1550$-$564~\citep{2014MNRAS.439L..65M}, MAXI~J1820+070~\citep{2021MNRAS.508.3104B}, MAXI~J1348$-$630~\citep{2022MNRAS.514.2839A}, and XTE~J1859+226~\citep{2022MNRAS.517.1469M}.

From the perspective of the radiative properties (rms and phase lag) of the bump, \citet{2003A&A...407.1039P} investigated the long-term variability of Cygnus~X-1 and proposed that the presence of the bump in the LHS originates in the hot accreting corona, which is responsible for most of the emission in the $\sim$2--13~keV in that state. For \grs, \citet{2001ApJ...558..276T} and \citet{2022MNRAS.514.2891Z} used X-ray and radio data in the HIMS to study the relation between the accretion-ejection scenario and the presence of either the bump or the radio jet. However, both \grs\ and Cygnus~X-1 are long-lasting black-hole sources~\footnote{Note that since 2018 \grs\ has been in a state of extreme low X-ray flux~\citep{2018ATel11828....1N}. It has been suggested that this low X-ray flux state of GRS 1915+105 is not similar to that of a typical X-ray transient, but is due to a super-Eddington accretion with a high local absorption near the black hole~\citep{2020A&A...639A..13K,2021MNRAS.503..152M} associated sometimes with a disk wind~\citep{2020ApJ...902..152N,2020ApJ...904...30M}.}.

Before this work, the bump in the typical black hole transient \gx\ had only been reported in a few \rxte\ observations~\citep{2000MNRAS.318..361N,2002ApJ...572..392B,2005A&A...440..207B}. From a systematic study of all the \rxte\ observations of \gx, as shown in Fig.~\ref{fig:pds}, we find that during the 2002, 2004, 2007 and 2010 outbursts, the bump is detected in 39 observations, all of which also show a type-C QPO in the LHS and HIMS. The bump is not present in all the observations, with the rms amplitude of the bump showing an evolution of the strength during the outburst. We generally lose the bump when the QPO moves to high frequencies, where the measurements of the rms amplitude of the bump decreases below the detection threshold. This could result from several reasons: As the QPO frequency increases, the frequency of the bump also increases~\citep[see, e.g.][]{2002ApJ...572..392B}. If the bump frequency becomes larger than 200~Hz, near the 256~Hz boundary of the PDS, the zero centered Lorentzian naturally fails to constrain the bump frequency. Another explanation of the nondetection of the bump can be that as \gx\ evolves from the LHS to the HIMS, the PDS is flatter and there are less time variability components~\citep[e.g.][]{2005A&A...440..207B}.

\subsection{The evolution of the type-C QPO in \gx}

In a previous study~\citep{2022MNRAS.514.2891Z}, we conducted a systematic study of the high-frequency bump in \grs, a black hole long-lasting X-ray transient prior to 2018~\citep{2021MNRAS.503..152M}. Fig.~2 in~\citet{2022MNRAS.514.2891Z} (see also Fig.~1 in~\citealt{2022NatAs...6..577M}) illustrates the HR versus type-C QPO frequency plot, which displays a spread of approximately 0.6 in HR at a specific QPO frequency, much larger than the errors. In the case of \gx, the points in Fig.~\ref{fig:rms} show much less scattering, with HR varying by approximately 0.2 at a specific QPO frequency. By comparing the rms amplitude of the bump in the HR versus the type-C QPO frequency plot in this paper with that in~\citet{2022MNRAS.514.2891Z}, it is apparent that Fig.~\ref{fig:rms} for \gx\ the data do not extend to the bottom left and top right parts as shown in Fig.~2 in~\citet{2022MNRAS.514.2891Z} (regions 5, 6, 13 to 16, and 19 to 24). These regions in Fig.~2 of~\citet{2022MNRAS.514.2891Z} are different from Fig.~\ref{fig:rms} of \gx: In regions 13 to 16, and 19 to 24, the rms of the bump is all upper limit and below 2\% in the bottom left regions; In regions 5 and 6, the bump is significantly detected.
The difference may be due to the fact that, with the appearance of type-C QPOs, \grs\ evolves only within a long-lasting HIMS, while \gx\ evolves through the LHS and HIMS, and undergoes state transitions from the HIMS to the SIMS.

Studies on black-hole X-ray transients have shown that the frequency of the type-C QPO increases as the source evolves from the LHS to the HIMS~\citep[e.g.][]{2011MNRAS.418.2292M,2020A&A...640A..18M,2021MNRAS.508..287S}. At the same time, the relative contribution of the hard component to the total X-ray flux decreases and the soft component becomes dominant~\citep{2010LNP...794...17G}. The anti-correlation between the hardness ratio and the type-C QPO frequency in Fig.~\ref{fig:rms} can also illustrate this point. A recent simulation based on the JED-SAD hybrid model naturally quantitatively explained the evolution of the type-C QPO frequency (Fig.~\ref{fig:rms_radio}) during the four outbursts of \gx. In this model, the spectral evolution of a transient XRB is driven by an interplay between a standard accretion disk (SAD) in the outer parts and a jet-emitting disk (JED) in the inner parts~\citep{2006A&A...447..813F,2020A&A...640A..18M}. By reconstructing the mass accretion rate and the flux contribution from the thermal disk and the corona, the type-C QPO frequency can be directly linked to the Keplerian frequency by a scaling factor of 100 at the transition radius, $r_{J}$, between the JED and SAD flows~\citep{2018A&A...615A..57M,2018A&A...617A..46M,2020A&A...640A..18M}. The type-C QPO frequency naturally evolves as the transition radius moves due to the JED-SAD magnetized structure~\citep[see Figures~2 and~3 in][]{2020A&A...640A..18M}. In Fig.~\ref{fig:rms_radio}, not only the frequency of the type-C QPO but also the flux density of the compact jet evolves during the LHS and HIMS of \gx, where there is a hint that in the rising phase of the outburst the change of the QPO frequency lags that of the jet flux density. The evolution of the radio flux can also be recovered from the JED-SAD model by introducing a scaling factor, $\Tilde{f}_{R}$~\citep{2020A&A...640A..18M,2022A&A...659A.194M}. Besides the temporal evolution of the type-C QPO and the radio flux density, the accretion-ejection structure in the JED-SAD model also provides an explanation for the complex multi-wavelength correlation in BH XRBs~\citep[e.g.][]{2003MNRAS.344...60G,2012MNRAS.423..590G,2013MNRAS.428.2500C}.

\subsection{The corona-jet radiative coupling}

When we detect it in \gx, the rms amplitude of the bump is always larger than $\sim$4--5\% (Fig.~\ref{fig:correlation}). The values correspond to the highest values of the rms amplitude of the bump, low radio flux and high corona temperature in \grs~\citep{2022NatAs...6..577M,2022MNRAS.514.2891Z}. As discussed first in~\citet{2001ApJ...558..276T} and then~\citet{2022NatAs...6..577M} and~\citet{2022MNRAS.514.2891Z}, the rms amplitude of the bump is anti-correlated with the radio flux and correlated with the corona temperature, indicating that the bump originates from the hot corona. \citet{2022MNRAS.514.2891Z} further proposed that under such condition, the exchange of accretion energy between the corona and jet is efficient. For the X-ray transient \gx\ in the LHS and the HIMS, the quasi-simultaneous X-ray and radio observations show that the bump is only detected when the radio flux is low and the corona temperature is high. When the temperature of the corona is higher than $\sim$30~keV, we mostly measure the upper limits of the rms of the bump. Note that these upper limits are within the error range of the significant measurements, such that the data are still consistent with the high corona temperature-high rms of the bump regime. Considering the findings in \grs, the presence of the bump in \gx\ suggests that in the hard state of this source, most of the accretion energy is directed to the corona instead of being used to eject the radio jet.


The weak corona-jet radiative coupling in the LHS of \gx\ is supported by the results of~\citet{2022A&A...659A.194M} using the JED-SAD model. Fig.~5 in~\citet{2022A&A...659A.194M} shows the JED-SAD transition radius, $r_{J}$, versus the mass accretion rate, $\dot{m}$, along with the energy transfer efficiency of the disk accretion energy in \gx. In the LHS, the energy transfer efficiency of advection through the corona is the strongest (up to 50\%), while that of jet and accretion flow cooling is below $\sim$25\%, justifying that the exchange of accretion energy to the jet through the corona is inefficient. Note, however, that the jet power is a parameter in the JED-SAD model, and that the model itself makes strong, although realistic, assumptions (see section 7.1 in~\citealt{2018A&A...615A..57M} and sections 3.1, 4.1, and 4.3 in~\citealt{2019A&A...626A.115M}).

The efficiency of the radiative coupling between different accretion-ejection components can be assessed through the empirical relationships involving the X-ray and radio fluxes in the hard state~\citep{2003MNRAS.344...60G,2003MNRAS.345.1057M}. We note that, however, this correlation could be biased by the selection of the X-ray energy range and the relative contribution from the soft disk and hard corona components, as well as any changes in radiative efficiencies of either the disk, the corona, or the jets during the outburst. Two branches can be identified within the X-ray--radio correlation: the normal branch and the outlier branch~\citep{2012MNRAS.423..590G}. In the outlier branch the source H~1743$-$322 exemplifies an efficient accretion-ejection coupling~\citep{2011MNRAS.414..677C,2020MNRAS.491L..29W}, whereas in the normal branch \gx\ shows an inefficient accretion-ejection coupling, characterized by an X-ray--radio correlation $L_{\text{radio}}\propto L^{0.6}_{X}$~\citep{2013MNRAS.428.2500C}. In the failed-transition outburst of \gx~\citep{2021MNRAS.507.5507A}, the correlation becomes even flatter, with $L_{\text{radio}}\propto L^{0.39}_{X}$, indicating in the hard-only outburst the coupling between the accretion flow and the jet is more inefficient~\citep{2021MNRAS.502..521D}.

Apart from the corona-jet radiative coupling, it is promising to further study the corona-jet morphological coupling by applying a time-dependent Comptonization model like \texttt{vKompth} to the radiative properties of the bump~\citep{2020MNRAS.492.1399K,2022MNRAS.515.2099B}. This model has been successful in explaining the radiative properties (rms and lags) of QPOs and deducing the size and the geometry of the corona~\citep[e.g.][]{2021MNRAS.501.3173G,2022NatAs...6..577M,2023MNRAS.520.5144Z}. The challenge we used to face for this is to measure the lags of the bump over a relatively broad frequency range. A novel method proposed by M{\'e}ndez et al.\ (2023, submitted) holds a promise to allow us to measure the lags of the bump that can be fitted by the time-dependent Comptonization model.

\nocite{mendez}




\section*{Acknowledgements}

We thank the referee for their constructive comments that help to improve the quality of this paper.
We fondly remember our colleague and dear friend, Tomaso M.\ Belloni, who sadly passed away on 26 August 2023, and who made crucial contributions to the field of X-ray timing over the past few decades.
YZ acknowledges support from the China Scholarship Council (CSC), no.\ 201906100030, and the Dutch Research Council (NWO) Rubicon Fellowship, file no.\ 019.231EN.021. MM acknowledges support from the research programme Athena with project number 184.034.002, which is (partly) financed by the Dutch Research Council (NWO). AAZ acknowledges support from the Polish National Science Center under the grant 2019/35/B/ST9/03944. FG is a CONICET researcher. FG acknowledges support by PIP 0113 (CONICET) and PIBAA 1275 (CONICET). This work received financial support from PICT2017-2865 (ANPCyT). DA acknowledges support from the Royal Society. TMB acknowledges financial contribution from PRIN INAF 2019 n.15. GM thanks the European Research Council (ERC) for support under the European Union's Horizon 2020 research and innovation programme (grant 834203). MM, AAZ, FG, TMB and GM thank the Team Meeting at the International Space Science Institute (Bern) for fruitful discussions, and were supported by the ISSI International Team project \#486.


\section*{Data Availability}

The X-ray data used in this article are accessible at NASA's High Energy Astrophysics Science Archive Research Center~\url{https://heasarc.gsfc.nasa.gov/}. The software GHATS for Fourier timing analysis is available at~\url{http://www.brera.inaf.it/utenti/belloni/GHATS_Package/Home.html}.



\bibliographystyle{mnras}
\bibliography{GX339_reference} 

\begin{thebibliography}{}
\makeatletter
\relax
\def\mn@urlcharsother{\let\do\@makeother \do\$\do\&\do\#\do\^\do\_\do\%\do\~}
\def\mn@doi{\begingroup\mn@urlcharsother \@ifnextchar [ {\mn@doi@}
  {\mn@doi@[]}}
\def\mn@doi@[#1]#2{\def\@tempa{#1}\ifx\@tempa\@empty \href
  {http://dx.doi.org/#2} {doi:#2}\else \href {http://dx.doi.org/#2} {#1}\fi
  \endgroup}
\def\mn@eprint#1#2{\mn@eprint@#1:#2::\@nil}
\def\mn@eprint@arXiv#1{\href {http://arxiv.org/abs/#1} {{\tt arXiv:#1}}}
\def\mn@eprint@dblp#1{\href {http://dblp.uni-trier.de/rec/bibtex/#1.xml}
  {dblp:#1}}
\def\mn@eprint@#1:#2:#3:#4\@nil{\def\@tempa {#1}\def\@tempb {#2}\def\@tempc
  {#3}\ifx \@tempc \@empty \let \@tempc \@tempb \let \@tempb \@tempa \fi \ifx
  \@tempb \@empty \def\@tempb {arXiv}\fi \@ifundefined
  {mn@eprint@\@tempb}{\@tempb:\@tempc}{\expandafter \expandafter \csname
  mn@eprint@\@tempb\endcsname \expandafter{\@tempc}}}

\bibitem[\protect\citeauthoryear{{Alabarta} et~al.,}{{Alabarta}
  et~al.}{2021}]{2021MNRAS.507.5507A}
{Alabarta} K.,  et~al., 2021, \mn@doi [\mnras] {10.1093/mnras/stab2241}, \href
  {https://ui.adsabs.harvard.edu/abs/2021MNRAS.507.5507A} {507, 5507}

\bibitem[\protect\citeauthoryear{{Alabarta}, {M{\'e}ndez}, {Garc{\'\i}a},
  {Peirano}, {Altamirano}, {Zhang}  \& {Karpouzas}}{{Alabarta}
  et~al.}{2022}]{2022MNRAS.514.2839A}
{Alabarta} K.,  {M{\'e}ndez} M.,  {Garc{\'\i}a} F.,  {Peirano} V.,
  {Altamirano} D.,  {Zhang} L.,   {Karpouzas} K.,  2022, \mn@doi [\mnras]
  {10.1093/mnras/stac1533}, \href
  {https://ui.adsabs.harvard.edu/abs/2022MNRAS.514.2839A} {514, 2839}

\bibitem[\protect\citeauthoryear{{Altamirano}, {van der Klis}, {M{\'e}ndez},
  {Jonker}, {Klein-Wolt}  \& {Lewin}}{{Altamirano}
  et~al.}{2008}]{2008ApJ...685..436A}
{Altamirano} D.,  {van der Klis} M.,  {M{\'e}ndez} M.,  {Jonker} P.~G.,
  {Klein-Wolt} M.,   {Lewin} W.~H.~G.,  2008, \mn@doi [\apj] {10.1086/590897},
  \href {https://ui.adsabs.harvard.edu/abs/2008ApJ...685..436A} {685, 436}

\bibitem[\protect\citeauthoryear{{Arnaud}}{{Arnaud}}{1996}]{1996ASPC..101...17A}
{Arnaud} K.~A.,  1996, in {Jacoby} G.~H.,  {Barnes} J.,  eds,  Astronomical
  Society of the Pacific Conference Series Vol. 101, Astronomical Data Analysis
  Software and Systems V. p.~17

\bibitem[\protect\citeauthoryear{{Bahramian} \& {Degenaar}}{{Bahramian} \&
  {Degenaar}}{2022}]{2022arXiv220610053B}
{Bahramian} A.,  {Degenaar} N.,  2022, \mn@doi [arXiv e-prints]
  {10.48550/arXiv.2206.10053}, \href
  {https://ui.adsabs.harvard.edu/abs/2022arXiv220610053B} {p. arXiv:2206.10053}

\bibitem[\protect\citeauthoryear{{Bellavita}, {Garc{\'\i}a}, {M{\'e}ndez}  \&
  {Karpouzas}}{{Bellavita} et~al.}{2022}]{2022MNRAS.515.2099B}
{Bellavita} C.,  {Garc{\'\i}a} F.,  {M{\'e}ndez} M.,   {Karpouzas} K.,  2022,
  \mn@doi [\mnras] {10.1093/mnras/stac1922}, \href
  {https://ui.adsabs.harvard.edu/abs/2022MNRAS.515.2099B} {515, 2099}

\bibitem[\protect\citeauthoryear{{Belloni} \& {Hasinger}}{{Belloni} \&
  {Hasinger}}{1990}]{1990A&A...230..103B}
{Belloni} T.,  {Hasinger} G.,  1990, \aap, \href
  {https://ui.adsabs.harvard.edu/abs/1990A&A...230..103B} {230, 103}

\bibitem[\protect\citeauthoryear{{Belloni}, {Psaltis}  \& {van der
  Klis}}{{Belloni} et~al.}{2002}]{2002ApJ...572..392B}
{Belloni} T.,  {Psaltis} D.,   {van der Klis} M.,  2002, \mn@doi [\apj]
  {10.1086/340290}, \href
  {https://ui.adsabs.harvard.edu/abs/2002ApJ...572..392B} {572, 392}

\bibitem[\protect\citeauthoryear{{Belloni}, {Homan}, {Casella}, {van der Klis},
  {Nespoli}, {Lewin}, {Miller}  \& {M{\'e}ndez}}{{Belloni}
  et~al.}{2005}]{2005A&A...440..207B}
{Belloni} T.,  {Homan} J.,  {Casella} P.,  {van der Klis} M.,  {Nespoli} E.,
  {Lewin} W.~H.~G.,  {Miller} J.~M.,   {M{\'e}ndez} M.,  2005, \mn@doi [\aap]
  {10.1051/0004-6361:20042457}, \href
  {https://ui.adsabs.harvard.edu/abs/2005A&A...440..207B} {440, 207}

\bibitem[\protect\citeauthoryear{{Belloni}, {Sanna}  \& {M{\'e}ndez}}{{Belloni}
  et~al.}{2012}]{2012MNRAS.426.1701B}
{Belloni} T.~M.,  {Sanna} A.,   {M{\'e}ndez} M.,  2012, \mn@doi [\mnras]
  {10.1111/j.1365-2966.2012.21634.x}, \href
  {https://ui.adsabs.harvard.edu/abs/2012MNRAS.426.1701B} {426, 1701}

\bibitem[\protect\citeauthoryear{{Bhargava}, {Belloni}, {Bhattacharya}, {Motta}
   \& {Ponti.}}{{Bhargava} et~al.}{2021}]{2021MNRAS.508.3104B}
{Bhargava} Y.,  {Belloni} T.,  {Bhattacharya} D.,  {Motta} S.,   {Ponti.} G.,
  2021, \mn@doi [\mnras] {10.1093/mnras/stab2848}, \href
  {https://ui.adsabs.harvard.edu/abs/2021MNRAS.508.3104B} {508, 3104}

\bibitem[\protect\citeauthoryear{{Bradt}, {Rothschild}  \& {Swank}}{{Bradt}
  et~al.}{1993}]{1993A&AS...97..355B}
{Bradt} H.~V.,  {Rothschild} R.~E.,   {Swank} J.~H.,  1993, \aaps, \href
  {https://ui.adsabs.harvard.edu/abs/1993A&AS...97..355B} {97, 355}

\bibitem[\protect\citeauthoryear{{Casella}, {Belloni}, {Homan}  \&
  {Stella}}{{Casella} et~al.}{2004}]{2004A&A...426..587C}
{Casella} P.,  {Belloni} T.,  {Homan} J.,   {Stella} L.,  2004, \mn@doi [\aap]
  {10.1051/0004-6361:20041231}, \href
  {https://ui.adsabs.harvard.edu/abs/2004A&A...426..587C} {426, 587}

\bibitem[\protect\citeauthoryear{{Casella}, {Belloni}  \& {Stella}}{{Casella}
  et~al.}{2005}]{2005ApJ...629..403C}
{Casella} P.,  {Belloni} T.,   {Stella} L.,  2005, \mn@doi [\apj]
  {10.1086/431174}, \href
  {https://ui.adsabs.harvard.edu/abs/2005ApJ...629..403C} {629, 403}

\bibitem[\protect\citeauthoryear{{Clavel}, {Rodriguez}, {Corbel}  \&
  {Coriat}}{{Clavel} et~al.}{2016}]{2016AN....337..435C}
{Clavel} M.,  {Rodriguez} J.,  {Corbel} S.,   {Coriat} M.,  2016, \mn@doi
  [Astronomische Nachrichten] {10.1002/asna.201612326}, \href
  {https://ui.adsabs.harvard.edu/abs/2016AN....337..435C} {337, 435}

\bibitem[\protect\citeauthoryear{{Corbel} et~al.,}{{Corbel}
  et~al.}{2001}]{2001ApJ...554...43C}
{Corbel} S.,  et~al., 2001, \mn@doi [\apj] {10.1086/321364}, \href
  {https://ui.adsabs.harvard.edu/abs/2001ApJ...554...43C} {554, 43}

\bibitem[\protect\citeauthoryear{{Corbel}, {Fender}, {Tzioumis}, {Tomsick},
  {Orosz}, {Miller}, {Wijnands}  \& {Kaaret}}{{Corbel}
  et~al.}{2002}]{2002Sci...298..196C}
{Corbel} S.,  {Fender} R.~P.,  {Tzioumis} A.~K.,  {Tomsick} J.~A.,  {Orosz}
  J.~A.,  {Miller} J.~M.,  {Wijnands} R.,   {Kaaret} P.,  2002, \mn@doi
  [Science] {10.1126/science.1075857}, \href
  {https://ui.adsabs.harvard.edu/abs/2002Sci...298..196C} {298, 196}

\bibitem[\protect\citeauthoryear{{Corbel}, {Coriat}, {Brocksopp}, {Tzioumis},
  {Fender}, {Tomsick}, {Buxton}  \& {Bailyn}}{{Corbel}
  et~al.}{2013}]{2013MNRAS.428.2500C}
{Corbel} S.,  {Coriat} M.,  {Brocksopp} C.,  {Tzioumis} A.~K.,  {Fender} R.~P.,
   {Tomsick} J.~A.,  {Buxton} M.~M.,   {Bailyn} C.~D.,  2013, \mn@doi [\mnras]
  {10.1093/mnras/sts215}, \href
  {https://ui.adsabs.harvard.edu/abs/2013MNRAS.428.2500C} {428, 2500}

\bibitem[\protect\citeauthoryear{{Coriat} et~al.,}{{Coriat}
  et~al.}{2011}]{2011MNRAS.414..677C}
{Coriat} M.,  et~al., 2011, \mn@doi [\mnras]
  {10.1111/j.1365-2966.2011.18433.x}, \href
  {https://ui.adsabs.harvard.edu/abs/2011MNRAS.414..677C} {414, 677}

\bibitem[\protect\citeauthoryear{{Corral-Santana}, {Casares},
  {Mu{\~n}oz-Darias}, {Bauer}, {Mart{\'\i}nez-Pais}  \&
  {Russell}}{{Corral-Santana} et~al.}{2016}]{2016A&A...587A..61C}
{Corral-Santana} J.~M.,  {Casares} J.,  {Mu{\~n}oz-Darias} T.,  {Bauer} F.~E.,
  {Mart{\'\i}nez-Pais} I.~G.,   {Russell} D.~M.,  2016, \mn@doi [\aap]
  {10.1051/0004-6361/201527130}, \href
  {https://ui.adsabs.harvard.edu/abs/2016A&A...587A..61C} {587, A61}

\bibitem[\protect\citeauthoryear{{Fabian}, {Rees}, {Stella}  \&
  {White}}{{Fabian} et~al.}{1989}]{1989MNRAS.238..729F}
{Fabian} A.~C.,  {Rees} M.~J.,  {Stella} L.,   {White} N.~E.,  1989, \mn@doi
  [\mnras] {10.1093/mnras/238.3.729}, \href
  {https://ui.adsabs.harvard.edu/abs/1989MNRAS.238..729F} {238, 729}

\bibitem[\protect\citeauthoryear{{Fender} \& {Belloni}}{{Fender} \&
  {Belloni}}{2004}]{2004ARA&A..42..317F}
{Fender} R.,  {Belloni} T.,  2004, \mn@doi [\araa]
  {10.1146/annurev.astro.42.053102.134031}, \href
  {https://ui.adsabs.harvard.edu/abs/2004ARA&A..42..317F} {42, 317}

\bibitem[\protect\citeauthoryear{{Fender}, {Garrington}, {McKay}, {Muxlow},
  {Pooley}, {Spencer}, {Stirling}  \& {Waltman}}{{Fender}
  et~al.}{1999}]{1999MNRAS.304..865F}
{Fender} R.~P.,  {Garrington} S.~T.,  {McKay} D.~J.,  {Muxlow} T.~W.~B.,
  {Pooley} G.~G.,  {Spencer} R.~E.,  {Stirling} A.~M.,   {Waltman} E.~B.,
  1999, \mn@doi [\mnras] {10.1046/j.1365-8711.1999.02364.x}, \href
  {https://ui.adsabs.harvard.edu/abs/1999MNRAS.304..865F} {304, 865}

\bibitem[\protect\citeauthoryear{{Fender}, {Belloni}  \& {Gallo}}{{Fender}
  et~al.}{2004}]{2004MNRAS.355.1105F}
{Fender} R.~P.,  {Belloni} T.~M.,   {Gallo} E.,  2004, \mn@doi [\mnras]
  {10.1111/j.1365-2966.2004.08384.x}, \href
  {https://ui.adsabs.harvard.edu/abs/2004MNRAS.355.1105F} {355, 1105}

\bibitem[\protect\citeauthoryear{{Fender}, {Belloni}  \& {Gallo}}{{Fender}
  et~al.}{2005}]{2005Ap&SS.300....1F}
{Fender} R.,  {Belloni} T.,   {Gallo} E.,  2005, \mn@doi [\apss]
  {10.1007/s10509-005-1201-z}, \href
  {https://ui.adsabs.harvard.edu/abs/2005Ap&SS.300....1F} {300, 1}

\bibitem[\protect\citeauthoryear{{Ferreira}, {Petrucci}, {Henri}, {Saug{\'e}}
  \& {Pelletier}}{{Ferreira} et~al.}{2006}]{2006A&A...447..813F}
{Ferreira} J.,  {Petrucci} P.~O.,  {Henri} G.,  {Saug{\'e}} L.,   {Pelletier}
  G.,  2006, \mn@doi [\aap] {10.1051/0004-6361:20052689}, \href
  {https://ui.adsabs.harvard.edu/abs/2006A&A...447..813F} {447, 813}

\bibitem[\protect\citeauthoryear{{Gallo}, {Fender}  \& {Pooley}}{{Gallo}
  et~al.}{2003}]{2003MNRAS.344...60G}
{Gallo} E.,  {Fender} R.~P.,   {Pooley} G.~G.,  2003, \mn@doi [\mnras]
  {10.1046/j.1365-8711.2003.06791.x}, \href
  {https://ui.adsabs.harvard.edu/abs/2003MNRAS.344...60G} {344, 60}

\bibitem[\protect\citeauthoryear{{Gallo}, {Miller}  \& {Fender}}{{Gallo}
  et~al.}{2012}]{2012MNRAS.423..590G}
{Gallo} E.,  {Miller} B.~P.,   {Fender} R.,  2012, \mn@doi [\mnras]
  {10.1111/j.1365-2966.2012.20899.x}, \href
  {https://ui.adsabs.harvard.edu/abs/2012MNRAS.423..590G} {423, 590}

\bibitem[\protect\citeauthoryear{{Garc{\'\i}a} et~al.,}{{Garc{\'\i}a}
  et~al.}{2014}]{2014ApJ...782...76G}
{Garc{\'\i}a} J.,  et~al., 2014, \mn@doi [\apj] {10.1088/0004-637X/782/2/76},
  \href {https://ui.adsabs.harvard.edu/abs/2014ApJ...782...76G} {782, 76}

\bibitem[\protect\citeauthoryear{{Garc{\'\i}a}, {Kallman}, {Bautista},
  {Mendoza}, {Deprince}, {Palmeri}  \& {Quinet}}{{Garc{\'\i}a}
  et~al.}{2018}]{2018ASPC..515..282G}
{Garc{\'\i}a} J.~A.,  {Kallman} T.~R.,  {Bautista} M.,  {Mendoza} C.,
  {Deprince} J.,  {Palmeri} P.,   {Quinet} P.,  2018, in Workshop on
  Astrophysical Opacities. p.~282 (\mn@eprint {arXiv} {1805.00581}),
  \mn@doi{10.48550/arXiv.1805.00581}

\bibitem[\protect\citeauthoryear{{Garc{\'\i}a}, {M{\'e}ndez}, {Karpouzas},
  {Belloni}, {Zhang}  \& {Altamirano}}{{Garc{\'\i}a}
  et~al.}{2021}]{2021MNRAS.501.3173G}
{Garc{\'\i}a} F.,  {M{\'e}ndez} M.,  {Karpouzas} K.,  {Belloni} T.,  {Zhang}
  L.,   {Altamirano} D.,  2021, \mn@doi [\mnras] {10.1093/mnras/staa3944},
  \href {https://ui.adsabs.harvard.edu/abs/2021MNRAS.501.3173G} {501, 3173}

\bibitem[\protect\citeauthoryear{{Garc{\'\i}a}, {Karpouzas}, {M{\'e}ndez},
  {Zhang}, {Zhang}, {Belloni}  \& {Altamirano}}{{Garc{\'\i}a}
  et~al.}{2022}]{2022MNRAS.513.4196G}
{Garc{\'\i}a} F.,  {Karpouzas} K.,  {M{\'e}ndez} M.,  {Zhang} L.,  {Zhang} Y.,
  {Belloni} T.,   {Altamirano} D.,  2022, \mn@doi [\mnras]
  {10.1093/mnras/stac1202}, \href
  {https://ui.adsabs.harvard.edu/abs/2022MNRAS.513.4196G} {513, 4196}

\bibitem[\protect\citeauthoryear{{Gilfanov}}{{Gilfanov}}{2010}]{2010LNP...794...17G}
{Gilfanov} M.,  2010, in {Belloni} T.,  ed., , Vol.~794, Lecture Notes in
  Physics, Berlin Springer Verlag.
p.~17, \mn@doi{10.1007/978-3-540-76937-8_2}

\bibitem[\protect\citeauthoryear{{\VAN{Haas}{de}{de} Haas}
  et~al.,}{{\VAN{Haas}{de}{de} Haas} et~al.}{2021}]{2021MNRAS.502..521D}
{\VAN{Haas}{de}{de} Haas} S.~E.~M.,  et~al., 2021, \mn@doi [\mnras]
  {10.1093/mnras/staa3853}, \href
  {https://ui.adsabs.harvard.edu/abs/2021MNRAS.502..521D} {502, 521}

\bibitem[\protect\citeauthoryear{{Heida}, {Jonker}, {Torres}  \&
  {Chiavassa}}{{Heida} et~al.}{2017}]{2017ApJ...846..132H}
{Heida} M.,  {Jonker} P.~G.,  {Torres} M.~A.~P.,   {Chiavassa} A.,  2017,
  \mn@doi [\apj] {10.3847/1538-4357/aa85df}, \href
  {https://ui.adsabs.harvard.edu/abs/2017ApJ...846..132H} {846, 132}

\bibitem[\protect\citeauthoryear{{Homan} \& {Belloni}}{{Homan} \&
  {Belloni}}{2005}]{2005Ap&SS.300..107H}
{Homan} J.,  {Belloni} T.,  2005, \mn@doi [\apss] {10.1007/s10509-005-1197-4},
  \href {https://ui.adsabs.harvard.edu/abs/2005Ap&SS.300..107H} {300, 107}

\bibitem[\protect\citeauthoryear{{Hynes} et~al.,}{{Hynes}
  et~al.}{2003}]{2003MNRAS.345..292H}
{Hynes} R.~I.,  et~al., 2003, \mn@doi [\mnras]
  {10.1046/j.1365-8711.2003.06938.x}, \href
  {https://ui.adsabs.harvard.edu/abs/2003MNRAS.345..292H} {345, 292}

\bibitem[\protect\citeauthoryear{{Hynes}, {Steeghs}, {Casares}, {Charles}  \&
  {O'Brien}}{{Hynes} et~al.}{2004}]{2004ApJ...609..317H}
{Hynes} R.~I.,  {Steeghs} D.,  {Casares} J.,  {Charles} P.~A.,   {O'Brien} K.,
  2004, \mn@doi [\apj] {10.1086/421014}, \href
  {https://ui.adsabs.harvard.edu/abs/2004ApJ...609..317H} {609, 317}

\bibitem[\protect\citeauthoryear{{Ingram} \& {Motta}}{{Ingram} \&
  {Motta}}{2019}]{2019NewAR..8501524I}
{Ingram} A.~R.,  {Motta} S.~E.,  2019, \mn@doi [\nar]
  {10.1016/j.newar.2020.101524}, \href
  {https://ui.adsabs.harvard.edu/abs/2019NewAR..8501524I} {85, 101524}

\bibitem[\protect\citeauthoryear{{Islam} \& {Zdziarski}}{{Islam} \&
  {Zdziarski}}{2018}]{2018MNRAS.481.4513I}
{Islam} N.,  {Zdziarski} A.~A.,  2018, \mn@doi [\mnras]
  {10.1093/mnras/sty2597}, \href
  {https://ui.adsabs.harvard.edu/abs/2018MNRAS.481.4513I} {481, 4513}

\bibitem[\protect\citeauthoryear{{Jahoda}, {Markwardt}, {Radeva}, {Rots},
  {Stark}, {Swank}, {Strohmayer}  \& {Zhang}}{{Jahoda}
  et~al.}{2006}]{2006ApJS..163..401J}
{Jahoda} K.,  {Markwardt} C.~B.,  {Radeva} Y.,  {Rots} A.~H.,  {Stark} M.~J.,
  {Swank} J.~H.,  {Strohmayer} T.~E.,   {Zhang} W.,  2006, \mn@doi [\apjs]
  {10.1086/500659}, \href
  {https://ui.adsabs.harvard.edu/abs/2006ApJS..163..401J} {163, 401}

\bibitem[\protect\citeauthoryear{{Kalemci}, {Tomsick}, {Rothschild},
  {Pottschmidt}, {Corbel}, {Wijnands}, {Miller}  \& {Kaaret}}{{Kalemci}
  et~al.}{2003}]{2003ApJ...586..419K}
{Kalemci} E.,  {Tomsick} J.~A.,  {Rothschild} R.~E.,  {Pottschmidt} K.,
  {Corbel} S.,  {Wijnands} R.,  {Miller} J.~M.,   {Kaaret} P.,  2003, \mn@doi
  [\apj] {10.1086/367693}, \href
  {https://ui.adsabs.harvard.edu/abs/2003ApJ...586..419K} {586, 419}

\bibitem[\protect\citeauthoryear{{Karpouzas}, {M{\'e}ndez}, {Ribeiro},
  {Altamirano}, {Blaes}  \& {Garc{\'\i}a}}{{Karpouzas}
  et~al.}{2020}]{2020MNRAS.492.1399K}
{Karpouzas} K.,  {M{\'e}ndez} M.,  {Ribeiro} E.~M.,  {Altamirano} D.,  {Blaes}
  O.,   {Garc{\'\i}a} F.,  2020, \mn@doi [\mnras] {10.1093/mnras/stz3502},
  \href {https://ui.adsabs.harvard.edu/abs/2020MNRAS.492.1399K} {492, 1399}

\bibitem[\protect\citeauthoryear{{Koljonen} \& {Tomsick}}{{Koljonen} \&
  {Tomsick}}{2020}]{2020A&A...639A..13K}
{Koljonen} K.~I.~I.,  {Tomsick} J.~A.,  2020, \mn@doi [\aap]
  {10.1051/0004-6361/202037882}, \href
  {https://ui.adsabs.harvard.edu/abs/2020A&A...639A..13K} {639, A13}

\bibitem[\protect\citeauthoryear{{Kubota} \& {Makishima}}{{Kubota} \&
  {Makishima}}{2004}]{2004ApJ...601..428K}
{Kubota} A.,  {Makishima} K.,  2004, \mn@doi [\apj] {10.1086/380433}, \href
  {https://ui.adsabs.harvard.edu/abs/2004ApJ...601..428K} {601, 428}

\bibitem[\protect\citeauthoryear{{Lightman} \& {White}}{{Lightman} \&
  {White}}{1988}]{1988ApJ...335...57L}
{Lightman} A.~P.,  {White} T.~R.,  1988, \mn@doi [\apj] {10.1086/166905}, \href
  {https://ui.adsabs.harvard.edu/abs/1988ApJ...335...57L} {335, 57}

\bibitem[\protect\citeauthoryear{{Liu}, {Jiang}, {Zhang}, {Bambi}, {Ji}, {Kong}
   \& {Zhang}}{{Liu} et~al.}{2022}]{2022MNRAS.513.4308L}
{Liu} H.,  {Jiang} J.,  {Zhang} Z.,  {Bambi} C.,  {Ji} L.,  {Kong} L.,
  {Zhang} S.,  2022, \mn@doi [\mnras] {10.1093/mnras/stac1178}, \href
  {https://ui.adsabs.harvard.edu/abs/2022MNRAS.513.4308L} {513, 4308}

\bibitem[\protect\citeauthoryear{{Liu} et~al.,}{{Liu}
  et~al.}{2023}]{2023ApJ...950....5L}
{Liu} H.,  et~al., 2023, \mn@doi [\apj] {10.3847/1538-4357/acca17}, \href
  {https://ui.adsabs.harvard.edu/abs/2023ApJ...950....5L} {950, 5}

\bibitem[\protect\citeauthoryear{{Marcel} et~al.,}{{Marcel}
  et~al.}{2018a}]{2018A&A...615A..57M}
{Marcel} G.,  et~al., 2018a, \mn@doi [\aap] {10.1051/0004-6361/201732069},
  \href {https://ui.adsabs.harvard.edu/abs/2018A&A...615A..57M} {615, A57}

\bibitem[\protect\citeauthoryear{{Marcel} et~al.,}{{Marcel}
  et~al.}{2018b}]{2018A&A...617A..46M}
{Marcel} G.,  et~al., 2018b, \mn@doi [\aap] {10.1051/0004-6361/201833124},
  \href {https://ui.adsabs.harvard.edu/abs/2018A&A...617A..46M} {617, A46}

\bibitem[\protect\citeauthoryear{{Marcel} et~al.,}{{Marcel}
  et~al.}{2019}]{2019A&A...626A.115M}
{Marcel} G.,  et~al., 2019, \mn@doi [\aap] {10.1051/0004-6361/201935060}, \href
  {https://ui.adsabs.harvard.edu/abs/2019A&A...626A.115M} {626, A115}

\bibitem[\protect\citeauthoryear{{Marcel} et~al.,}{{Marcel}
  et~al.}{2020}]{2020A&A...640A..18M}
{Marcel} G.,  et~al., 2020, \mn@doi [\aap] {10.1051/0004-6361/202037539}, \href
  {https://ui.adsabs.harvard.edu/abs/2020A&A...640A..18M} {640, A18}

\bibitem[\protect\citeauthoryear{{Marcel} et~al.,}{{Marcel}
  et~al.}{2022}]{2022A&A...659A.194M}
{Marcel} G.,  et~al., 2022, \mn@doi [\aap] {10.1051/0004-6361/202141375}, \href
  {https://ui.adsabs.harvard.edu/abs/2022A&A...659A.194M} {659, A194}

\bibitem[\protect\citeauthoryear{{Markert}, {Canizares}, {Clark}, {Lewin},
  {Schnopper}  \& {Sprott}}{{Markert} et~al.}{1973}]{1973ApJ...184L..67M}
{Markert} T.~H.,  {Canizares} C.~R.,  {Clark} G.~W.,  {Lewin} W.~H.~G.,
  {Schnopper} H.~W.,   {Sprott} G.~F.,  1973, \mn@doi [\apjl] {10.1086/181290},
  \href {https://ui.adsabs.harvard.edu/abs/1973ApJ...184L..67M} {184, L67}

\bibitem[\protect\citeauthoryear{{M{\'e}ndez} \& {et al.,}}{{M{\'e}ndez} \& {et
  al.,}}{2023}]{mendez}
{M{\'e}ndez} M.,  {et al.,} 2023, submitted to MNRAS

\bibitem[\protect\citeauthoryear{{M{\'e}ndez} \& {van der Klis}}{{M{\'e}ndez}
  \& {van der Klis}}{1997}]{1997ApJ...479..926M}
{M{\'e}ndez} M.,  {van der Klis} M.,  1997, \mn@doi [\apj] {10.1086/303914},
  \href {https://ui.adsabs.harvard.edu/abs/1997ApJ...479..926M} {479, 926}

\bibitem[\protect\citeauthoryear{{M{\'e}ndez}, {Altamirano}, {Belloni}  \&
  {Sanna}}{{M{\'e}ndez} et~al.}{2013}]{2013MNRAS.435.2132M}
{M{\'e}ndez} M.,  {Altamirano} D.,  {Belloni} T.,   {Sanna} A.,  2013, \mn@doi
  [\mnras] {10.1093/mnras/stt1431}, \href
  {https://ui.adsabs.harvard.edu/abs/2013MNRAS.435.2132M} {435, 2132}

\bibitem[\protect\citeauthoryear{{M{\'e}ndez}, {Karpouzas}, {Garc{\'\i}a},
  {Zhang}, {Zhang}, {Belloni}  \& {Altamirano}}{{M{\'e}ndez}
  et~al.}{2022}]{2022NatAs...6..577M}
{M{\'e}ndez} M.,  {Karpouzas} K.,  {Garc{\'\i}a} F.,  {Zhang} L.,  {Zhang} Y.,
  {Belloni} T.~M.,   {Altamirano} D.,  2022, \mn@doi [Nature Astronomy]
  {10.1038/s41550-022-01617-y}, \href
  {https://ui.adsabs.harvard.edu/abs/2022NatAs...6..577M} {6, 577}

\bibitem[\protect\citeauthoryear{{Merloni}, {Heinz}  \& {di Matteo}}{{Merloni}
  et~al.}{2003}]{2003MNRAS.345.1057M}
{Merloni} A.,  {Heinz} S.,   {di Matteo} T.,  2003, \mn@doi [\mnras]
  {10.1046/j.1365-2966.2003.07017.x}, \href
  {https://ui.adsabs.harvard.edu/abs/2003MNRAS.345.1057M} {345, 1057}

\bibitem[\protect\citeauthoryear{{Miller} et~al.,}{{Miller}
  et~al.}{2008}]{2008ApJ...679L.113M}
{Miller} J.~M.,  et~al., 2008, \mn@doi [\apjl] {10.1086/589446}, \href
  {https://ui.adsabs.harvard.edu/abs/2008ApJ...679L.113M} {679, L113}

\bibitem[\protect\citeauthoryear{{Miller} et~al.,}{{Miller}
  et~al.}{2020}]{2020ApJ...904...30M}
{Miller} J.~M.,  et~al., 2020, \mn@doi [\apj] {10.3847/1538-4357/abbb31}, \href
  {https://ui.adsabs.harvard.edu/abs/2020ApJ...904...30M} {904, 30}

\bibitem[\protect\citeauthoryear{{Mitsuda} et~al.,}{{Mitsuda}
  et~al.}{1984}]{1984PASJ...36..741M}
{Mitsuda} K.,  et~al., 1984, \pasj, \href
  {https://ui.adsabs.harvard.edu/abs/1984PASJ...36..741M} {36, 741}

\bibitem[\protect\citeauthoryear{{Motta}}{{Motta}}{2016}]{2016AN....337..398M}
{Motta} S.~E.,  2016, \mn@doi [Astronomische Nachrichten]
  {10.1002/asna.201612320}, \href
  {https://ui.adsabs.harvard.edu/abs/2016AN....337..398M} {337, 398}

\bibitem[\protect\citeauthoryear{{Motta} \& {Belloni}}{{Motta} \&
  {Belloni}}{2023}]{2023arXiv230700867M}
{Motta} S.~E.,  {Belloni} T.~M.,  2023, \mn@doi [arXiv e-prints]
  {10.48550/arXiv.2307.00867}, \href
  {https://ui.adsabs.harvard.edu/abs/2023arXiv230700867M} {p. arXiv:2307.00867}

\bibitem[\protect\citeauthoryear{{Motta}, {Mu{\~n}oz-Darias}, {Casella},
  {Belloni}  \& {Homan}}{{Motta} et~al.}{2011}]{2011MNRAS.418.2292M}
{Motta} S.,  {Mu{\~n}oz-Darias} T.,  {Casella} P.,  {Belloni} T.,   {Homan} J.,
   2011, \mn@doi [\mnras] {10.1111/j.1365-2966.2011.19566.x}, \href
  {https://ui.adsabs.harvard.edu/abs/2011MNRAS.418.2292M} {418, 2292}

\bibitem[\protect\citeauthoryear{{Motta}, {Homan}, {Mu{\~n}oz Darias},
  {Casella}, {Belloni}, {Hiemstra}  \& {M{\'e}ndez}}{{Motta}
  et~al.}{2012}]{2012MNRAS.427..595M}
{Motta} S.,  {Homan} J.,  {Mu{\~n}oz Darias} T.,  {Casella} P.,  {Belloni}
  T.~M.,  {Hiemstra} B.,   {M{\'e}ndez} M.,  2012, \mn@doi [\mnras]
  {10.1111/j.1365-2966.2012.22037.x}, \href
  {https://ui.adsabs.harvard.edu/abs/2012MNRAS.427..595M} {427, 595}

\bibitem[\protect\citeauthoryear{{Motta}, {Belloni}, {Stella},
  {Mu{\~n}oz-Darias}  \& {Fender}}{{Motta} et~al.}{2014a}]{2014MNRAS.437.2554M}
{Motta} S.~E.,  {Belloni} T.~M.,  {Stella} L.,  {Mu{\~n}oz-Darias} T.,
  {Fender} R.,  2014a, \mn@doi [\mnras] {10.1093/mnras/stt2068}, \href
  {https://ui.adsabs.harvard.edu/abs/2014MNRAS.437.2554M} {437, 2554}

\bibitem[\protect\citeauthoryear{{Motta}, {Munoz-Darias}, {Sanna}, {Fender},
  {Belloni}  \& {Stella}}{{Motta} et~al.}{2014b}]{2014MNRAS.439L..65M}
{Motta} S.~E.,  {Munoz-Darias} T.,  {Sanna} A.,  {Fender} R.,  {Belloni} T.,
  {Stella} L.,  2014b, \mn@doi [\mnras] {10.1093/mnrasl/slt181}, \href
  {https://ui.adsabs.harvard.edu/abs/2014MNRAS.439L..65M} {439, L65}

\bibitem[\protect\citeauthoryear{{Motta} et~al.,}{{Motta}
  et~al.}{2021}]{2021MNRAS.503..152M}
{Motta} S.~E.,  et~al., 2021, \mn@doi [\mnras] {10.1093/mnras/stab511}, \href
  {https://ui.adsabs.harvard.edu/abs/2021MNRAS.503..152M} {503, 152}

\bibitem[\protect\citeauthoryear{{Motta}, {Belloni}, {Stella}, {Pappas},
  {Casares}, {Mu{\~n}oz-Darias}, {Torres}  \& {Yanes-Rizo}}{{Motta}
  et~al.}{2022}]{2022MNRAS.517.1469M}
{Motta} S.~E.,  {Belloni} T.,  {Stella} L.,  {Pappas} G.,  {Casares} J.,
  {Mu{\~n}oz-Darias} A.~T.,  {Torres} M.~A.~P.,   {Yanes-Rizo} I.~V.,  2022,
  \mn@doi [\mnras] {10.1093/mnras/stac2142}, \href
  {https://ui.adsabs.harvard.edu/abs/2022MNRAS.517.1469M} {517, 1469}

\bibitem[\protect\citeauthoryear{{Negoro} et~al.,}{{Negoro}
  et~al.}{2018}]{2018ATel11828....1N}
{Negoro} H.,  et~al., 2018, The Astronomer's Telegram, \href
  {https://ui.adsabs.harvard.edu/abs/2018ATel11828....1N} {11828, 1}

\bibitem[\protect\citeauthoryear{{Neilsen}, {Homan}, {Steiner}, {Marcel},
  {Cackett}, {Remillard}  \& {Gendreau}}{{Neilsen}
  et~al.}{2020}]{2020ApJ...902..152N}
{Neilsen} J.,  {Homan} J.,  {Steiner} J.~F.,  {Marcel} G.,  {Cackett} E.,
  {Remillard} R.~A.,   {Gendreau} K.,  2020, \mn@doi [\apj]
  {10.3847/1538-4357/abb598}, \href
  {https://ui.adsabs.harvard.edu/abs/2020ApJ...902..152N} {902, 152}

\bibitem[\protect\citeauthoryear{{Nowak}}{{Nowak}}{2000}]{2000MNRAS.318..361N}
{Nowak} M.~A.,  2000, \mn@doi [\mnras] {10.1046/j.1365-8711.2000.03668.x},
  \href {https://ui.adsabs.harvard.edu/abs/2000MNRAS.318..361N} {318, 361}

\bibitem[\protect\citeauthoryear{{Parker} et~al.,}{{Parker}
  et~al.}{2016}]{2016ApJ...821L...6P}
{Parker} M.~L.,  et~al., 2016, \mn@doi [\apjl] {10.3847/2041-8205/821/1/L6},
  \href {https://ui.adsabs.harvard.edu/abs/2016ApJ...821L...6P} {821, L6}

\bibitem[\protect\citeauthoryear{{Pooley} \& {Fender}}{{Pooley} \&
  {Fender}}{1997}]{1997MNRAS.292..925P}
{Pooley} G.~G.,  {Fender} R.~P.,  1997, \mn@doi [\mnras]
  {10.1093/mnras/292.4.925}, \href
  {https://ui.adsabs.harvard.edu/abs/1997MNRAS.292..925P} {292, 925}

\bibitem[\protect\citeauthoryear{{Pottschmidt} et~al.,}{{Pottschmidt}
  et~al.}{2003}]{2003A&A...407.1039P}
{Pottschmidt} K.,  et~al., 2003, \mn@doi [\aap] {10.1051/0004-6361:20030906},
  \href {https://ui.adsabs.harvard.edu/abs/2003A&A...407.1039P} {407, 1039}

\bibitem[\protect\citeauthoryear{{Psaltis}, {Belloni}  \& {van der
  Klis}}{{Psaltis} et~al.}{1999}]{1999ApJ...520..262P}
{Psaltis} D.,  {Belloni} T.,   {van der Klis} M.,  1999, \mn@doi [\apj]
  {10.1086/307436}, \href
  {https://ui.adsabs.harvard.edu/abs/1999ApJ...520..262P} {520, 262}

\bibitem[\protect\citeauthoryear{{Reid}, {McClintock}, {Steiner}, {Steeghs},
  {Remillard}, {Dhawan}  \& {Narayan}}{{Reid}
  et~al.}{2014}]{2014ApJ...796....2R}
{Reid} M.~J.,  {McClintock} J.~E.,  {Steiner} J.~F.,  {Steeghs} D.,
  {Remillard} R.~A.,  {Dhawan} V.,   {Narayan} R.,  2014, \mn@doi [\apj]
  {10.1088/0004-637X/796/1/2}, \href
  {https://ui.adsabs.harvard.edu/abs/2014ApJ...796....2R} {796, 2}

\bibitem[\protect\citeauthoryear{{Remillard} \& {McClintock}}{{Remillard} \&
  {McClintock}}{2006}]{2006ARA&A..44...49R}
{Remillard} R.~A.,  {McClintock} J.~E.,  2006, \mn@doi [\araa]
  {10.1146/annurev.astro.44.051905.092532}, \href
  {https://ui.adsabs.harvard.edu/abs/2006ARA&A..44...49R} {44, 49}

\bibitem[\protect\citeauthoryear{{Rothschild} et~al.,}{{Rothschild}
  et~al.}{1998}]{1998ApJ...496..538R}
{Rothschild} R.~E.,  et~al., 1998, \mn@doi [\apj] {10.1086/305377}, \href
  {https://ui.adsabs.harvard.edu/abs/1998ApJ...496..538R} {496, 538}

\bibitem[\protect\citeauthoryear{{Russell}, {Miller-Jones}, {Maccarone},
  {Yang}, {Fender}  \& {Lewis}}{{Russell} et~al.}{2011}]{2011ApJ...739L..19R}
{Russell} D.~M.,  {Miller-Jones} J.~C.~A.,  {Maccarone} T.~J.,  {Yang} Y.~J.,
  {Fender} R.~P.,   {Lewis} F.,  2011, \mn@doi [\apjl]
  {10.1088/2041-8205/739/1/L19}, \href
  {https://ui.adsabs.harvard.edu/abs/2011ApJ...739L..19R} {739, L19}

\bibitem[\protect\citeauthoryear{{Shui} et~al.,}{{Shui}
  et~al.}{2021}]{2021MNRAS.508..287S}
{Shui} Q.~C.,  et~al., 2021, \mn@doi [\mnras] {10.1093/mnras/stab2521}, \href
  {https://ui.adsabs.harvard.edu/abs/2021MNRAS.508..287S} {508, 287}

\bibitem[\protect\citeauthoryear{{Tetarenko}, {Casella}, {Miller-Jones},
  {Sivakoff}, {Tetarenko}, {Maccarone}, {Gandhi}  \& {Eikenberry}}{{Tetarenko}
  et~al.}{2019}]{2019MNRAS.484.2987T}
{Tetarenko} A.~J.,  {Casella} P.,  {Miller-Jones} J.~C.~A.,  {Sivakoff} G.~R.,
  {Tetarenko} B.~E.,  {Maccarone} T.~J.,  {Gandhi} P.,   {Eikenberry} S.,
  2019, \mn@doi [\mnras] {10.1093/mnras/stz165}, \href
  {https://ui.adsabs.harvard.edu/abs/2019MNRAS.484.2987T} {484, 2987}

\bibitem[\protect\citeauthoryear{{Trudolyubov}}{{Trudolyubov}}{2001}]{2001ApJ...558..276T}
{Trudolyubov} S.~P.,  2001, \mn@doi [\apj] {10.1086/322466}, \href
  {https://ui.adsabs.harvard.edu/abs/2001ApJ...558..276T} {558, 276}

\bibitem[\protect\citeauthoryear{{Verner}, {Ferland}, {Korista}  \&
  {Yakovlev}}{{Verner} et~al.}{1996}]{1996ApJ...465..487V}
{Verner} D.~A.,  {Ferland} G.~J.,  {Korista} K.~T.,   {Yakovlev} D.~G.,  1996,
  \mn@doi [\apj] {10.1086/177435}, \href
  {https://ui.adsabs.harvard.edu/abs/1996ApJ...465..487V} {465, 487}

\bibitem[\protect\citeauthoryear{{Vincentelli} et~al.,}{{Vincentelli}
  et~al.}{2021}]{2021MNRAS.503..614V}
{Vincentelli} F.~M.,  et~al., 2021, \mn@doi [\mnras] {10.1093/mnras/stab475},
  \href {https://ui.adsabs.harvard.edu/abs/2021MNRAS.503..614V} {503, 614}

\bibitem[\protect\citeauthoryear{{Wang-Ji} et~al.,}{{Wang-Ji}
  et~al.}{2018}]{2018ApJ...855...61W}
{Wang-Ji} J.,  et~al., 2018, \mn@doi [\apj] {10.3847/1538-4357/aaa974}, \href
  {https://ui.adsabs.harvard.edu/abs/2018ApJ...855...61W} {855, 61}

\bibitem[\protect\citeauthoryear{{Wijnands} \& {van der Klis}}{{Wijnands} \&
  {van der Klis}}{1999}]{1999ApJ...514..939W}
{Wijnands} R.,  {van der Klis} M.,  1999, \mn@doi [\apj] {10.1086/306993},
  \href {https://ui.adsabs.harvard.edu/abs/1999ApJ...514..939W} {514, 939}

\bibitem[\protect\citeauthoryear{{Williams} et~al.,}{{Williams}
  et~al.}{2020}]{2020MNRAS.491L..29W}
{Williams} D.~R.~A.,  et~al., 2020, \mn@doi [\mnras] {10.1093/mnrasl/slz152},
  \href {https://ui.adsabs.harvard.edu/abs/2020MNRAS.491L..29W} {491, L29}

\bibitem[\protect\citeauthoryear{{Wilms}, {Allen}  \& {McCray}}{{Wilms}
  et~al.}{2000}]{2000ApJ...542..914W}
{Wilms} J.,  {Allen} A.,   {McCray} R.,  2000, \mn@doi [\apj] {10.1086/317016},
  \href {https://ui.adsabs.harvard.edu/abs/2000ApJ...542..914W} {542, 914}

\bibitem[\protect\citeauthoryear{{Yan} \& {Yu}}{{Yan} \&
  {Yu}}{2015}]{2015ApJ...805...87Y}
{Yan} Z.,  {Yu} W.,  2015, \mn@doi [\apj] {10.1088/0004-637X/805/2/87}, \href
  {https://ui.adsabs.harvard.edu/abs/2015ApJ...805...87Y} {805, 87}

\bibitem[\protect\citeauthoryear{{Yang} et~al.,}{{Yang}
  et~al.}{2023}]{2023MNRAS.tmp..760Y}
{Yang} Z.-X.,  et~al., 2023, \mn@doi [\mnras] {10.1093/mnras/stad795}, \href
  {https://ui.adsabs.harvard.edu/abs/2023MNRAS.tmp..760Y} {}

\bibitem[\protect\citeauthoryear{{Zdziarski}}{{Zdziarski}}{2014}]{2014MNRAS.444.1113Z}
{Zdziarski} A.~A.,  2014, \mn@doi [\mnras] {10.1093/mnras/stu1525}, \href
  {https://ui.adsabs.harvard.edu/abs/2014MNRAS.444.1113Z} {444, 1113}

\bibitem[\protect\citeauthoryear{{Zdziarski}, {Johnson}  \&
  {Magdziarz}}{{Zdziarski} et~al.}{1996}]{1996MNRAS.283..193Z}
{Zdziarski} A.~A.,  {Johnson} W.~N.,   {Magdziarz} P.,  1996, \mn@doi [\mnras]
  {10.1093/mnras/283.1.193}, \href
  {https://ui.adsabs.harvard.edu/abs/1996MNRAS.283..193Z} {283, 193}

\bibitem[\protect\citeauthoryear{{Zdziarski}, {Zi{\'o}{\l}kowski}  \&
  {Miko{\l}ajewska}}{{Zdziarski} et~al.}{2019}]{2019MNRAS.488.1026Z}
{Zdziarski} A.~A.,  {Zi{\'o}{\l}kowski} J.,   {Miko{\l}ajewska} J.,  2019,
  \mn@doi [\mnras] {10.1093/mnras/stz1787}, \href
  {https://ui.adsabs.harvard.edu/abs/2019MNRAS.488.1026Z} {488, 1026}

\bibitem[\protect\citeauthoryear{{Zhang}, {Jahoda}, {Swank}, {Morgan}  \&
  {Giles}}{{Zhang} et~al.}{1995}]{1995ApJ...449..930Z}
{Zhang} W.,  {Jahoda} K.,  {Swank} J.~H.,  {Morgan} E.~H.,   {Giles} A.~B.,
  1995, \mn@doi [\apj] {10.1086/176111}, \href
  {https://ui.adsabs.harvard.edu/abs/1995ApJ...449..930Z} {449, 930}

\bibitem[\protect\citeauthoryear{{Zhang}, {M{\'e}ndez}, {Garc{\'\i}a},
  {Karpouzas}, {Zhang}, {Liu}, {Belloni}  \& {Altamirano}}{{Zhang}
  et~al.}{2022}]{2022MNRAS.514.2891Z}
{Zhang} Y.,  {M{\'e}ndez} M.,  {Garc{\'\i}a} F.,  {Karpouzas} K.,  {Zhang} L.,
  {Liu} H.,  {Belloni} T.~M.,   {Altamirano} D.,  2022, \mn@doi [\mnras]
  {10.1093/mnras/stac1050}, \href
  {https://ui.adsabs.harvard.edu/abs/2022MNRAS.514.2891Z} {514, 2891}

\bibitem[\protect\citeauthoryear{{Zhang} et~al.,}{{Zhang}
  et~al.}{2023}]{2023MNRAS.520.5144Z}
{Zhang} Y.,  et~al., 2023, \mn@doi [\mnras] {10.1093/mnras/stad460}, \href
  {https://ui.adsabs.harvard.edu/abs/2023MNRAS.520.5144Z} {520, 5144}

\bibitem[\protect\citeauthoryear{{{\.Z}ycki}, {Done}  \& {Smith}}{{{\.Z}ycki}
  et~al.}{1999}]{1999MNRAS.309..561Z}
{{\.Z}ycki} P.~T.,  {Done} C.,   {Smith} D.~A.,  1999, \mn@doi [\mnras]
  {10.1046/j.1365-8711.1999.02885.x}, \href
  {https://ui.adsabs.harvard.edu/abs/1999MNRAS.309..561Z} {309, 561}

\makeatother
\end{thebibliography}



\appendix

\section{HID of \gx}

\begin{figure*}
    \includegraphics[width=0.99\textwidth]{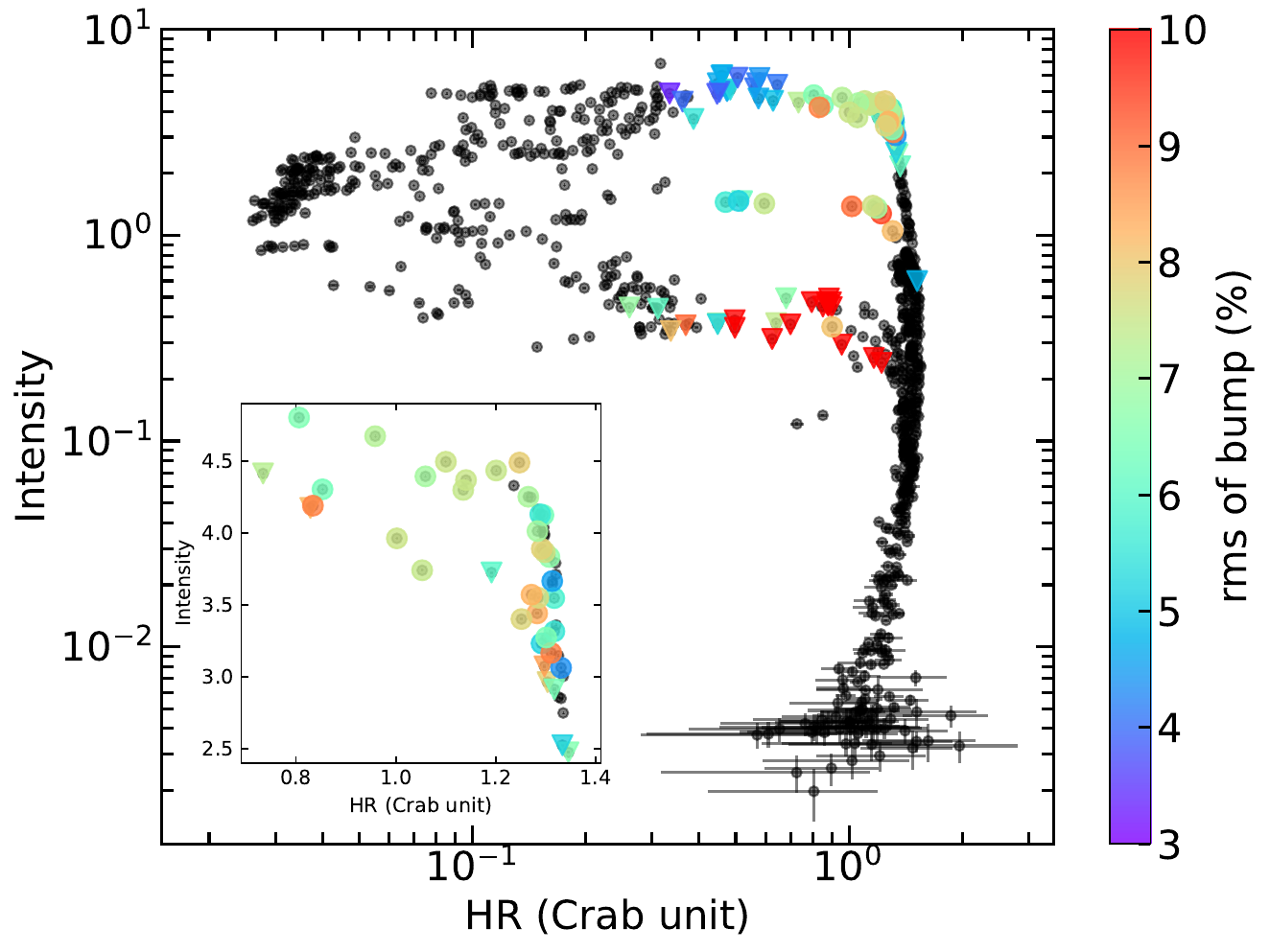}
    \caption{The HID of \gx\ using \rxte\ archival data. The colored points in the first horizontal branch correspond to the rising phases of the 2002, 2007, and 2010 outbursts; The colored points in the second horizontal branch represent the rising phase of the 2004 outburst, while those in the third horizontal branch pertain to the decaying phases of the 2002, 2004, 2007, and 2010 outbursts. Colored circles denote measurements of the rms amplitude of the bump, while down triangles correspond the upper limits of the rms amplitude of the bump. The small panel inside provides an enlarged view of the top-right corner. Note that in the decaying phase, the red triangles represent the upper limits of the rms that exceed 10\%.}
    \label{fig:hid}
\end{figure*}

Fig.~\ref{fig:hid} illustrates the HID of \gx\ as observed. The colored points represent observations during the appearance of type-C QPOs and the rms amplitude of the bump during the 2002, 2004, 2007, and 2010 outbursts.

In the top two horizontal branches, which are also in the rising phase, we observe an initial increase in the rms amplitude of the bump from below 4\% to approximately 9\% during the LHS-to-HIMS transition. Subsequently, it decreases to below 4\% during the HIMS-to-SIMS transition. In the bottom horizontal branch, which is in the decaying phase, we generally find upper limits for the rms amplitude of the bump. It is noteworthy that, even though we set the upper boundary of the color bar to be 10\%, during the decaying phase, the upper limits of the rms amplitude of the bump can reach as high as 30\% (see also the right panel of Fig.~\ref{fig:correlation}). This can be attributed to the power spectra in the decaying phase, which are very noisy with large error bars at high frequencies.

\bsp	
\label{lastpage}
\end{document}